\documentclass[11pt, a4paper]{article}
\usepackage{jheppub}
\usepackage{amsfonts}
\usepackage{amsmath}
\usepackage{url}
\usepackage{eurosym}
\usepackage{braket}
\usepackage{graphicx}
\usepackage{dsfont}
\usepackage{slashed}
\usepackage[T1]{fontenc}
\usepackage[utf8]{inputenc}
\usepackage{BOONDOX-cal}

\def\bbone{{\mathchoice {\rm 1\mskip-4mu l} {\rm 1\mskip-4mu l}
{\rm 1\mskip-4.5mu l} {\rm 1\mskip-5mu l}}}
\def\SO{\mathsf{SO}}

\def\OSp{\mathsf{OSp}}

\DeclareFontFamily{U}{MnSymbolC}{}
\DeclareSymbolFont{MnSyC}{U}{MnSymbolC}{m}{n}
\DeclareFontShape{U}{MnSymbolC}{m}{n}{
    <-6>  MnSymbolC5
   <6-7>  MnSymbolC6
   <7-8>  MnSymbolC7
   <8-9>  MnSymbolC8
   <9-10> MnSymbolC9
  <10-12> MnSymbolC10
  <12->   MnSymbolC12}{}
\DeclareMathSymbol{\intprod}{\mathbin}{MnSyC}{'270}

\begin{document}

\title{Charges of supergravity}

\author[a]{Remigiusz Durka}

\author[a,b]{Jerzy Kowalski-Glikman}
\author[b]{Rene Payne}
\affiliation[a]{University of Wroc\l{}aw, Faculty of Physics and Astronomy, pl.\ M.\ Borna 9, 50-204 Wroc\l{}aw, Poland}
\affiliation[b]{National Centre for Nuclear Research, Pasteura 7, 02-093 Warsaw, Poland}
\emailAdd{remigiusz.durka@uwr.edu.pl}
\emailAdd{jerzy.kowalski-glikman@uwr.edu.pl}
\emailAdd{rene.payne@ncbj.gov.pl}

%%%%%%%%%%%%%%%%%%%%%%%%%%%%%%%%%%%%%%%%%%%%%%%%%%%%%%%%%%%%%%%%%%

\abstract{We study conserved charges of $\mathcal{N}=1$ supergravity formulated as a constrained BF theory based on the $\OSp(1|4)$ superalgebra. Using the covariant phase space formalism, we derive bulk and boundary contributions to the symplectic structure and construct charges associated with Lorentz transformations, supersymmetry, translations, and diffeomorphisms. We show that the algebra of boundary charges reproduces the expected superalgebra, while translational charges vanish on-shell due to the super-torsion constraint, leaving Lorentz and supersymmetry as the non-trivial generators.}

%%%%%%%%%%%%%%%%%%%%%%%%%%%%%%%%%%%%%%%%%%%%%%%%%%%%%%%%%%%%%%%%%%

%\arxivnumber{1234.5678}

\maketitle

%%%%%%%%%%%%%%%%%%%%%%%%%%%%%%%%%%%%%%%%%%%%%%%%%%%%%%%%%%%%%%%%%%

\section{Introduction}

The idea of describing gravity as a gauge theory has attracted sustained interest over the years, as it offers a promising route toward reconciling General Relativity with the framework of quantum field theory. Among the various proposals, the MacDowell--Mansouri formalism \cite{MacDowell:1977jt} is particularly notable for its conceptual clarity and elegance. In this approach, gravity with a cosmological constant is reformulated as a gauge theory of the (Anti-)de Sitter group $\SO(2,3)$ or $\SO(1,4)$. One of its key features is that it unifies the tetrad and the spin connection into a single connection one-form, allowing spacetime symmetries to be treated in a fully gauge-theoretic manner.

A powerful and closely related perspective emerges when this construction is expressed in the language of BF theories. Although pure BF theory is topological and does not contain local degrees of freedom, it can be appropriately modified — either through constraints or deformations — to reproduce General Relativity \cite{Plebanski:1977zz, Smolin:2003qu, Freidel:2005ak}. This formulation is especially appealing because of its deep connections with spin foam models and Loop Quantum Gravity. Moreover, it naturally accommodates the Barbero–Immirzi parameter, which plays an important role in quantum theory \cite{Holst:1995pc, Durka:2011yv}. Extending this framework further to include supersymmetry leads to supergravity, where one gains, among other things, a more systematic and robust definition of conserved charges.

In earlier work \cite{Durka:2009pf} (see also \cite{Eder:2021rgt}), it was demonstrated that $\mathcal{N}=1$ supergravity with a negative cosmological constant can be formulated as a constrained BF theory based on the superalgebra $\OSp(1|4)$. This construction can be viewed as a supersymmetric extension of the Freidel--Smolin--Starodubtsev model \cite{Smolin:2003qu, Freidel:2005ak}, and it reproduces the standard supergravity action supplemented by additional topological terms, such as the Euler, Pontryagin and Nieh--Yan, and the Holst contribution.

More recently, there has been renewed interest in the role of boundaries in gravitational theories, particularly in light of the “corner symmetry” proposal \cite{Donnelly:2016auv,Freidel:2020xyx, Freidel:2020svx, Freidel:2020ayo} (for reviews, see, e.g., \cite{Ciambelli:2022vot,Assanioussi:2023jyq,Speziale:2025lkm}). The central idea behind this approach is that, in the presence of boundaries, gauge redundancies cease to be mere redundancies: instead, they give rise to genuine physical symmetries acting on the boundary phase space. The surface charges associated with these symmetries form an algebra that captures important dynamical and thermodynamical properties of spacetime, including, for instance, aspects related to black hole entropy. While this framework has been successfully developed for purely bosonic gravity \cite{Durka:2021ftc}, its extension to supersymmetric theories remains largely unexplored and constitutes an important open problem.

In this paper, we take a step in this direction by systematically deriving the conserved charges of $\OSp(1|4)$ supergravity formulated as a constrained BF theory. Our main objective is to better understand the symplectic structure of the theory in the presence of boundaries and to compute the algebra of charges associated with local Lorentz transformations, diffeomorphisms, and supersymmetry.

The structure of the paper is as follows. In Section 2, we review the formulation of supergravity as a constrained BF theory. We introduce the $\OSp(1|4)$ connection and its curvature, construct the action including the Immirzi parameter, and derive the equations of motion for both the fundamental fields and the auxiliary $\mathbb{B}$ fields. Section 3 is devoted to a detailed analysis of the symmetries of the theory, including the transformation laws of the component fields under supersymmetry, Lorentz transformations, and diffeomorphisms. In Sections 4 and 5, we apply the covariant phase space formalism to derive the symplectic potential, carefully separating bulk and corner contributions, and thereby identifying the relevant boundary charges. Finally, in Section 6, we compute the algebra of these corner charges and explicitly demonstrate that the Poisson brackets of the supercharges close, reproducing the expected superalgebra at the boundary.

%%%%%%%%%%%%%%%%%%%%%%%%%%%%%%%%%%%%%%%%%%%%%%%%%%%%%%%%%%%%%%%%%%

\section{Supergravity as a constrained BF theory}

It is well known that gravity can be formulated as a gauge theory of the Poincar\'e group in flat spacetime and of $\SO(2,3)$ or $\SO(1,4)$ in the presence of a negative or positive cosmological constant, respectively. Similarly, supergravity \cite{VanNieuwenhuizen:1981ae} with negative cosmological constant \cite{Townsend:1977qa} can be understood as a gauge theory of the superalgebra $\OSp(1|4)$. The defining commutators and anticommutators of this graded algebra are presented in Appendix~\ref{AppendixA}.

In this section, we recall the results of \cite{Durka:2009pf} exploring a supersymmetric version of the BF model proposed by Freidel and Starodubtsev \cite{Freidel:2005ak}. This construction has its roots in the work of Plebanski \cite{Plebanski:1977zz} and the scheme used in the MacDowell--Mansouri model \cite{MacDowell:1977jt}. In what follows, we use definitions, sign conventions, constants, gammas, etc., as in \cite{Durka:2012wd} and we recall them in Appendix~\ref{AppendixB}.

%%%%%%%%%%%%%%%%%%%%%%%%%%%%%%%%%%%%%%%%%%%%%%%%%%%%%%%%%%%%%%%%%%

\subsection{Connection, covariant derivative, and curvature}

The gauge field $\mathbb{A}$ is a one-form valued in a Lie algebra or superalgebra. In the case at hand, when the gauge (super)algebra is $\OSp(1|4)$ (see Appendix~\ref{AppendixA} for this algebra (anti)commutators and Appendix~\ref{AppendixB} for conventions concerning Dirac $\gamma$-matrices), the gauge field decomposes into the Lorentz connection $\omega^{ab}$ associated with Lorentz generators, the tetrad $e^a$ associated with translations, and the gravitino $\psi$ associated with the supersymmetry generators. Because canonically the gauge field (one-form) is dimensionless while the dimension of gravitino is $[-1/2]$ we introduce the compensating constant $\kappa$ of dimension $[1/2]$ defined by
\begin{equation}\label{kappa}
\kappa^2=\frac{4\pi G}{\ell}\,,
\end{equation}
to make the dimensions right and get the factor 1/2 in front of the gravitino kinetic term. In this formula $G$ is Newton's constant, and $\ell$ is a length scale related to the cosmological constant $-\frac{1}{\ell^2}=\frac{\Lambda}{3}$. Similarly, with the help of $\ell$, we also assure the correct dimension in the translational part of the connection, the tetrad.

The gauge field \cite{Durka:2012wd} is then written as
\begin{equation}\label{connection}
\mathbb{A}=\frac{1}{2}A^{IJ} M_{IJ}+\kappa\bar\psi^\alpha Q_\alpha=\frac{1}{2}\omega^{ab}M_{ab}+\frac{1}{\ell}e^aP_a+\kappa\bar\psi^\alpha Q_\alpha\,.
\end{equation}
Accordingly the (super)curvature 
\begin{align}\label{curvdef}
 \mathbb{F}= d\mathbb{A} -\frac{i}{2} [[\mathbb{A} \stackrel{\wedge}{,}\, \mathbb{A}]]\,,
\end{align}
splits into bosonic and fermionic parts
\begin{equation}
\mathbb{F}=\frac12\, F^{(s)}{}^{IJ}\, M_{IJ} + \bar{\mathcal F}^\alpha Q_\alpha=\frac12\, F^{(s)}{}^{ab}\, M_{ab}+ F^{(s)}{}^{a}\, P_{a}+ \bar{\mathcal F}^\alpha Q_\alpha\,.
\end{equation}
In the formula \eqref{curvdef} we denote by $[[\ast\stackrel{\wedge}{,}\,\ast]]$ the bracket of two $\OSp(1|4)$-valued forms. This bracket becomes anticommutator for two fermions and commutator otherwise.

The bosonic (super)curvatures associated with local Lorentz and translational symmetries are
\begin{align}\label{AdS_curvatures}
F^{(s)}{}^{ab} =F^{ab}-\frac{\kappa^2}{2}\,\bar\psi\wedge\gamma^{ab}\psi\,,\qquad F^{(s)a}\equiv F^{(s)a4}=F^{a}+\frac{\kappa^2}{2}\, \bar\psi\wedge\gamma^{a}\psi\,,
\end{align}
with the AdS curvature $F^{ab} =R^{ab}+\frac{1}{\ell^2} e^a\wedge e^b$ and the torsion $T^{a}=\ell F^{a}=D^\omega e^a$, where the Lorentzian covariant differential is
\begin{align}
D^\omega e^a=d e^a + \omega^{a}{}_{b}\wedge e^{b}\,.
\end{align}

The fermionic curvature $\mathcal F$ can be defined with the help of a covariant differential
\begin{align}
\mathcal D^A\bar{\psi}&= d\bar{\psi}-\frac{1}{4}\omega^{ab}\wedge\bar{\psi}\,\gamma_{ab}-\frac{1}{2\ell} e^{a}\wedge\bar{\psi}\,\gamma_{a}\,,\qquad
\mathcal D^A\psi= d\psi+\frac{1}{4}\omega^{ab}\wedge\gamma_{ab}\,\psi+\frac{1}{2\ell} e^{a}\wedge\gamma_{a}\,\psi\,,
\end{align}
where we identify
\begin{align}
\mathcal{D}^\omega\psi
&=
d\psi+\frac{1}{4}\omega^{ab}\wedge\gamma_{ab}\psi\,,
&
\mathcal{D}^\omega\bar{\psi}
&=
d\bar{\psi}-\frac{1}{4}\omega^{ab}\wedge\bar{\psi}\gamma_{ab}\,.
\end{align}
Then the fermionic curvatures are given by
\begin{align}\label{spinorial_curvature}
{\mathcal F} &=\kappa \mathcal D^A\psi=\kappa \left(\mathcal D^\omega\psi+\frac{1}{2\ell}e^a\wedge
\gamma_a\psi\right)\,,\\
\bar{\mathcal F} &=\kappa \mathcal D^A\bar{\psi}=\kappa \left(\mathcal D^\omega\bar{\psi}-\frac{1}{2\ell}e^a\wedge
\bar{\psi}\gamma_a\right)\,.
\end{align}
Below it will be convenient to use the complete covariant differential associated with connection \eqref{connection} that acts on gauge (super)algebra-valued $n$-forms
\begin{align}\label{gencovder}
\mathbb{D}^{\mathbb{A}}(\ast) = d(\ast) - i [[\mathbb{A}\stackrel{\wedge}{,} (\ast)]]\,.
\end{align}
One can check that
\begin{align}\label{gencurv}
\mathbb{F} =- i [[{\mathbb D}^{\mathbb{A}}\stackrel{\wedge}{,}{\mathbb D}^{\mathbb{A}}]]\,.
\end{align}
It follows directly from \eqref{gencovder} and \eqref{gencurv} that variation of the curvature can be expressed as a covariant derivative of variation of the gauge potential,
\begin{align}\label{genBianchi_2}
\delta\mathbb{F}(\mathbb{A}) ={\mathbb D}^{\mathbb{A}} \delta\mathbb{A}\,.
\end{align}
It follows also that the curvature satisfies Bianchi identity
\begin{align}\label{genBianchi_1}
\mathbb{D}^{\mathbb{A}} \mathbb{F} =0\,,
 \end{align}
 which decomposes into
\begin{align}\label{genBianchi_1decomposition}
D^{A} F^{(s)IJ} +\kappa\bar{\psi}\wedge \gamma^{IJ}\mathcal{F}=0\,,\\
\mathcal{D}^{A} \bar{\mathcal{F}} +\kappa \bar{\psi}\wedge\gamma_{IJ}F^{(s)IJ}=0\,.
\end{align}
Decomposing the connection into Lorentz, translational and supersymmetric parts (see \cite{Andrianopoli:2014aqa}) we can rewrite these equations as 
\begin{align}
& D^{\omega} F^{(s)ab} + \frac{1}{\ell}\, e^a\wedge F^{(s)b}-\frac{1}{\ell}\, e^b\wedge F^{(s)a}+\kappa\bar{\psi}\wedge \gamma^{ab}\mathcal{F}=0\,,\label{genBianchi_1decomposition_explicite1}\\
& D^{\omega} F^{(s)a} - \frac{1}{\ell}\, e_b\wedge F^{(s)ab}-\kappa\bar{\psi} \wedge\gamma^{a}\mathcal{F}=0\,,\label{genBianchi_1decomposition_explicite2}\\
& \mathcal{D}^{\omega} \bar{\mathcal{F}}-\frac{1}{2\ell}e^{a}\,\wedge\bar{\mathcal{F}}\,\gamma_{a} +\kappa \bar{\psi}\wedge\gamma_{ab}F^{(s)ab}+2\kappa \bar{\psi}\wedge\gamma_{a}F^{(s)a}=0\,.\label{genBianchi_1decomposition_explicite3}
\end{align}
Also note the useful identity $\mathcal{D}^{\omega}\mathcal{D}^{\omega}\psi=\frac{1}{4}R^{ab}(\omega)\wedge\gamma_{ab}\psi$.

With these definitions, we can now proceed to define the BF action for supergravity.

%%%%%%%%%%%%%%%%%%%%%%%%%%%%%%%%%%%%%%%%%%%%%%%%%%%%%%%%%%%%%%%%%%

\subsection{Super BF theory}

The bosonic constrained BF theory action proposed by Freidel, Smolin, and Starodubtsev \cite{Smolin:2003qu, Freidel:2005ak} 
\begin{equation}\label{action_BF}
 16\pi \, S({A},B)= \int F^{IJ}\wedge B_{IJ} -\frac{\beta}{2} B^{IJ}\wedge B_{IJ} - \frac{\alpha}{4}\epsilon^{abcd4} B_{ab}\wedge B_{cd}\,,
\end{equation}
where $B_{IJ}$ is the $\SO(2,3)$-algebra-valued 2-form, yields an extension of the MacDowell--Mansouri scheme \cite{MacDowell:1977jt} that includes the Barbero-Immirzi parameter \cite{Holst:1995pc, BarberoG:1994eia} (here expressed as $\gamma=\frac{\beta}{\alpha}$)\footnote{We realize that there is a danger of confusing the Barbero-Immirzi parameter with Dirac matrices. However all the Dirac matrices have some indices, and the $\gamma$ without an index is always understood to be the Barbero-Immirzi parameter.}. After solving field equations for $B$ fields with dimensionless parameters $\alpha$, $\beta$ related to the gravitational and cosmological constants
\begin{equation}\label{constants}
\alpha= \frac{G\Lambda}{3\,(1+\gamma^2)}\,,\qquad
\beta= \frac{\gamma G\Lambda}{3\,(1+\gamma^2)},\quad \mathrm{with}\quad \Lambda=-\frac3{\ell^2}\,,
\end{equation} 
we see that the action \eqref{action_BF} provides all six terms of the first-order gravity, namely the Einstein-Cartan action with cosmological term appended by Holst, Euler, Pontryagin, and Nieh--Yan terms
\begin{align}\label{action_behind_BF}
 32\pi G\, S&=\int R^{ab}\wedge e^{c}\wedge e^{d}\,\epsilon_{abcd}+
\frac{1}{\,2\ell^2}\int e^{a}\wedge e^{b}\wedge e^{c}\wedge e^{d}\, \epsilon_{abcd}\nonumber\\
&+\frac{2}{\gamma}\int R^{ab}\wedge e_{a}\wedge e_{b}+\frac{\ell^2}{2}\int R^{ab}\wedge R^{cd} \,\epsilon_{abcd}\nonumber\\
&-\ell^2\gamma\int R^{ab}\wedge R_{ab}+\frac{\gamma^2+1}{\gamma}\int 2\,(T^a \wedge T_a - R^{ab}\wedge e_a \wedge e_b)\,.
\end{align}
One sees that the field equations coming from this action are the Einstein equations. 

The local supersymmetric extension of the action \eqref{action_BF} was proposed in \cite{Durka:2009pf} (see also \cite{Eder:2021rgt}). Introducing the fermionic partner ${\mathcal B}$ of the bosonic field $B^{(s)IJ}$, we write the action as
\begin{align}
16\pi S(A,\psi, B, \cal B) &=\int \mathcal{L}^{sugra}_{bosonic}+ \mathcal{L}^{sugra}_{fermionic}\nonumber\\
&= \int B^{(s)IJ}\wedge F^{(s)}_{IJ} - \frac\beta2\, B^{(s)IJ}\wedge B_{IJ}^{(s)} -\frac\alpha4\, \epsilon_{abcd} \, B^{(s)ab}\wedge B^{(s)cd} \nonumber\\
&~~+4\, \int \bar{\mathcal B}\wedge {\mathcal F}- \frac\beta2\,\bar{\mathcal B}\wedge {\mathcal B}-\frac{\alpha}{4}\, \bar{\mathcal B} \wedge \gamma^5 {\mathcal B}\, ,\label{Lagrangian_sugra_fermionic}
\end{align}
which can be further expanded to curvature quantities (see Appendix~\ref{AppendixC}).

In what follows it will be convenient to write this action more compactly. To this end we introduce the two-form field $\mathbb B = (B^{(s)IJ}, \mathcal B)$ and its dual $\star\mathbb B= (\star B^{(s)ab}, \star B^{(s)a}, \star \mathcal B) =(\epsilon^{abcd}B^{(s)}_{cd}, 0, \gamma^5 \mathcal B)$ to write the action \eqref{Lagrangian_sugra_fermionic} as
\begin{align}
16\pi S(\mathbb{A}, \mathbb{B}) &= \int \left\langle\mathbb B \wedge \mathbb F \right\rangle- \frac\beta2\left\langle\mathbb B \wedge \mathbb B\right\rangle - \frac\alpha4 \left\langle\mathbb B \wedge \star\mathbb B\right\rangle\,,\label{LSUGRA}
\end{align}
with the inner product defined as
\begin{align}
\left\langle\mathbb B \wedge \mathbb F \right\rangle = B^{(s)IJ}\wedge F^{(s)}_{IJ}+4\bar{\mathcal B}\wedge {\mathcal F}\,.
\end{align}
The variation of the action is
\begin{align}\label{variation_full_BF_action}
16\pi\delta S &= \int \delta\mathbb B \wedge \left(\mathbb F - \beta \mathbb B - \frac\alpha2 \star\mathbb B \right) +\mathbb B \wedge \left(\mathbb D^{\mathbb A}\delta \mathbb A\right)\,.
\end{align}

The $\mathbb B$ field equations obtained from the first term in \eqref{variation_full_BF_action} can be decomposed into a bosonic part, which reads
\begin{align}\label{B_bosonic_field_equations}
B^{(s)a}&=\frac{1}{\beta} F^{(s)a} =\frac{1}{\beta}\left(F^{a}+\frac{\kappa^2}{2}\bar{\psi}\wedge\gamma^{a}\,\psi\right)\, ,\\
B^{(s)ab}&=\frac{1}{2(\alpha^2+\beta^2)} \left(\beta\delta^{ab}_{cd}
-\alpha\,\epsilon^{ab}{}_{cd}\right) F^{(s)cd}\nonumber\\
& =\frac{1}{2(\alpha^2+\beta^2)} \left(\beta\delta^{ab}_{cd}
-\alpha\,\epsilon^{ab}{}_{cd}\right)\left(F^{cd}-\frac{\kappa^2}{2}\bar{\psi}\wedge\gamma^{cd}\,\psi\right)\, ,
\end{align}
and its fermionic counterpart 
\begin{align}\label{B_fermionic_field_equations}
\mathcal{B}&=\frac{1}{\alpha^2+\beta^2}\left(\beta\bbone
-\alpha\,\gamma^5\right)\mathcal{F}\,,
\\
\bar{\mathcal{B}}
&=
\frac{1}{\alpha^2+\beta^2}
\bar{\mathcal{F}}\left(\beta\bbone-\alpha\,\gamma^5\right)\,.
\end{align}

Integrating by parts the second term in \eqref{variation_full_BF_action} and neglecting the boundary term for a moment we derive the gauge field $\mathbb A$ field equations
\begin{align}\label{Afieldeq}
\mathbb D^{\mathbb A}\mathbb B=0\,.
\end{align}
The field equation \eqref{Afieldeq} can be decomposed into fermionic and bosonic parts
\begin{align}
&\mathcal{D}^A\bar{\mathcal{B}}+\kappa \bar{\psi}\wedge\gamma^{IJ} B^{(s)}_{IJ}=0\,,\label{Bfield_I}\\
&D^A B^{(s)IJ}+\kappa \bar{\psi}\wedge\gamma^{IJ}\mathcal{B} = 0\,,\label{Bfield_II}
\end{align}
whose full decomposition is given in Appendix~\ref{AppendixC}.

Let us notice that as shown in \eqref{vanishingsupertorsion}, the equation $D^A B^{(s)ab}=0$ enforces the vanishing of the super-torsion
\begin{align}\label{super-torsion}
0=F^{(s)a}=\frac{1}{\ell}D^\omega e^a+\frac{\kappa^2}{2}\bar{\psi}\wedge\gamma^{a}\psi\,,
\end{align}
which can be rewritten as a relation between torsion and gravitino field
\begin{align}
T^{a} + 2\pi G \,\bar{\psi} \wedge\gamma^{a}\psi=0\,.
\end{align}

%%%%%%%%%%%%%%%%%%%%%%%%%%%%%%%%%%%%%%%%%%%%%%%%%%%%%%%%%%%%%%%%%%

\section{Symmetries}

The infinitesimal gauge transformations of the gauge field are defined in terms of the covariant derivative
\begin{equation}\label{c3-17}
\delta_\Upsilon \mathbb{A}_{\mu}=\partial_\mu \Upsilon
-i[\mathbb{A}_{\mu}, \Upsilon]\equiv \mathbb{D}^{\mathbb{A}}_\mu \Upsilon\, ,
\end{equation}
where the gauge parameter $\Upsilon$ decomposes into parameters of
local Lorentz, translation, and supercharge symmetries
\begin{equation}\label{c3-18}
\Upsilon=\frac{1}{2}\lambda^{IJ}M_{IJ}+\bar\epsilon^\alpha Q_\alpha=\frac{1}{2}\lambda^{ab}M_{ab}+\zeta^a P_a+\bar\epsilon^\alpha Q_\alpha\,.
\end{equation}

%%%%%%%%%%%%%%%%%%%%%%%%%%%%%%%%%%%%%%%%%%%%%%%%%%%%%%%%%%%%%%%%%%

\subsection{Supersymmetry}

Using the above formula one can immediately derive the supersymmetry transformations
\begin{equation}
\delta_\epsilon e^{a} = -\ell\kappa\,\bar\epsilon\, \gamma^{a}\,\psi\,,\qquad
\delta_\epsilon \omega^{ab} = \kappa\,\bar\epsilon\, \gamma^{ab}\,\psi\, ,\qquad
\delta_\epsilon \bar{\psi} = \frac{1}{\kappa}(\mathcal D^{\omega}\bar\epsilon-\frac{1}{2\ell}e^a\bar\epsilon\gamma_a) \, ,
\end{equation}
where $\mathcal{D}^\omega \bar\epsilon=d\bar\epsilon-\frac14\omega^{ab}\, \bar\epsilon\gamma_{ab}\,.$
The supersymmetry transformations of the curvatures can be easily obtained from
\begin{equation}\label{c3-19}
\delta \mathbb{F}_{\mu\nu} = \mathbb{D}_\mu \delta\mathbb{A}_{\nu} -\mathbb{D}_\nu \delta\mathbb{A}_{\mu} = [\mathbb{D}_\mu, \mathbb{D}_\nu]\Upsilon=i[\Upsilon,\mathbb{F}_{\mu\nu}]\,, 
\end{equation}
therefore we have
\begin{equation}\label{c3-20}
 \delta_\epsilon F_{}^{(s)a} = -\bar\epsilon \gamma^{a} {\mathcal F}_{}\,,\qquad
\delta_\epsilon F_{}^{(s)ab} = \bar\epsilon \gamma^{ab} {\mathcal F}_{}\,,\qquad
\delta_\epsilon \bar{\mathcal F}_{} = -\frac{1}{4}\bar\epsilon \gamma^{ab}F^{(s)}_{ab}-\frac{1}{2}\bar\epsilon\gamma_a F^{(s)a}\,.
\end{equation}

One can further show that the first two terms of the Lagrangian \eqref{LSUGRA} are invariant under local supersymmetry if the components of the field ${\mathbb B}=(B^{(s)}, {\mathcal B})$ transform as 
\begin{equation}\label{c3-23}
 \delta_\epsilon B_{}^{(s)a} = -\bar\epsilon \gamma^{a} {\mathcal B}_{}\,,\qquad
\delta_\epsilon B_{}^{(s)ab} = \bar\epsilon \gamma^{ab} {\mathcal B}_{}\,,\qquad
\delta_\epsilon \bar{\mathcal B}_{} = -\frac{1}{4}\bar\epsilon \gamma^{ab}B^{(s)}_{ab}-\frac{1}{2}\bar\epsilon\gamma_a B^{(s)a}\,.
\end{equation}
However, the gauge breaking term in \eqref{Lagrangian_sugra_fermionic}, 
\begin{equation}
-\frac\alpha4\, \epsilon_{abcd} \, B^{(s)ab}\wedge B^{(s)cd} -{\alpha}\, \bar{\mathcal B}\wedge\gamma^5 {\mathcal B}\,, \label{nonsusy} 
\end{equation}
is not invariant under supersymmetry transformations given in (\ref{c3-23}). Indeed its variation is 
\begin{equation}\label{nonsusy_var}
\alpha B^{(s)a}\wedge\bar\epsilon\gamma_a\gamma^5{\mathcal B}\,,
\end{equation}
and does not vanish. As it was shown in \cite{Durka:2009pf} the action obtained after solving the algebraic field equations for $B$ and $\mathcal{B}$ is nevertheless invariant under local supersymmetry when one employs the $1.5$ formalism. The very reason is that the term \eqref{nonsusy_var} is proportional to super-torsion, which vanishes as a result of the (algebraic) field equation for the Lorentz connection $\omega$. Since in this paper we are mostly interested in conserved charges, which are defined on-shell, the off-shell local supersymmetry non-invariance is not going to be relevant to what follows.

%%%%%%%%%%%%%%%%%%%%%%%%%%%%%%%%%%%%%%%%%%%%%%%%%%%%%%%%%%%%%%%%%%

\subsection{Translations and Lorentz transformations}

Before we go any further, for completeness we list the remaining transformations. The Lorentz transformations and translations are as follows:
\begin{align}
\delta_{\lambda}\omega^{ab}&= D^\omega \lambda^{ab}\,, \qquad
\frac{1}{\ell}\delta_{\lambda} e^{a}=-\frac{1}{\ell}\lambda^a_{\;b} \,e^b\,,\qquad \delta_{\lambda}\bar{\psi}=-\frac{1}{4}\bar{\psi}\gamma_{ab}\lambda^{ab}\,,\\
 \delta_{\zeta}\omega^{ab}&=\frac{1}{\ell}(e^a\, \zeta^b-e^b\, \zeta^a)\,,\qquad \frac{1}{\ell}\delta_{\zeta} e^{a}=D^\omega \zeta^a\,,\qquad \delta_{\zeta}\bar{\psi}=-\frac{1}{2}\bar{\psi}\gamma_{a}\zeta^{a}\,,
\end{align}
while for the components of curvature we find
\begin{align}
\delta_{\lambda} F^{(s)ab}=-[\lambda,F^{(s)}]^{ab} \,,\qquad \frac{1}{\ell}\delta_{\lambda } T^{(s)a}=-\frac{1}{\ell}\lambda^a_{\;b}T^{(s)b}\,,\\
\delta_{\zeta } F^{(s)ab}=-\frac{1}{\ell}[\zeta,T^{(s)}]^{ab} \,,\qquad \frac{1}{\ell}\delta_{\zeta} T^{(s)a}=\zeta_b\,F^{(s)ab}\,.
\end{align}
We add here the transformations of the $B$ fields:
\begin{equation}
\delta_\lambda B^{(s)ab}=-[\lambda,B^{(s)}]^{ab}\,,\quad \delta_\lambda{B}^{(s)a}=-\lambda^a{}_{b}{B}^{(s)b}\,,
\end{equation}
\begin{equation}
 \delta_\zeta B^{(s)ab}=({B}^{(s)a} \zeta^b-{B}^{(s)b} \zeta^a)\,, \quad \delta_\zeta
 {B}^{(s)a}=B^{(s)ab}\zeta_b\,,
\end{equation}
\begin{equation}
\delta_{\lambda}\bar{\mathcal{B}}= \frac{1}{4}\bar{\mathcal{B}}\gamma_{ab}\lambda^{ab}\,,\qquad \delta_{\zeta}\bar{\mathcal{B}}= \frac{1}{2}\bar{\mathcal{B}}\gamma_{a}\zeta^{a}\,.
\end{equation}

%%%%%%%%%%%%%%%%%%%%%%%%%%%%%%%%%%%%%%%%%%%%%%%%%%%%%%%%%%%%%%%%%%

\subsection{Diffeomorphisms}

We now turn to diffeomorphisms generated by a vector field $\xi^\mu$. The transformation laws are given by the action of the Lie derivative,
\begin{align}
 \mathcal{L}_{\xi}(\ast)=\xi\lrcorner(d\,\ast)+d(\xi\lrcorner\,\ast)\,,
\end{align}
and read
\begin{align}
\delta_{\xi} A_\mu^{IJ} &= \mathcal{L}_{\xi}A^{IJ}_{\mu}\,,\\
\delta_{\xi} \psi_\mu &= \mathcal{L}_{\xi}\psi_\mu\,,\\
\delta_{\xi} B_{\mu\nu}^{(s)IJ} &= \mathcal{L}_{\xi}B_{\mu\nu}^{(s)IJ}
= \xi^\rho \partial_\rho B_{\mu\nu}^{(s)IJ}
+ B_{\rho\nu}^{(s)IJ}\partial_\mu \xi^\rho
+ B_{\mu\rho}^{(s)IJ}\partial_\nu \xi^\rho\,,\\
\delta_{\xi} \mathcal{B}_{\mu\nu} &= \mathcal{L}_{\xi}\mathcal{B}_{\mu\nu}
= \xi^\rho \partial_\rho \mathcal{B}_{\mu\nu}
+ \mathcal{B}_{\rho\nu}\partial_\mu \xi^\rho
+ \mathcal{B}_{\mu\rho}\partial_\nu \xi^\rho\,.
\end{align}
The Lorentz connection, tetrad, and gravitino transform as follows
\begin{align}
\delta_{\xi} \omega_\mu^{ab} &= \xi^\nu\partial_\nu\omega_\mu^{ab}
+\omega_\nu^{ab}\partial_\mu \xi^{\nu}\,,\\
\delta_{\xi} e_\mu^{a} &= \xi^\nu\partial_\nu e_\mu^{a}
+e_\nu^{a}\partial_\mu \xi^\nu\,,\\
\delta_{\xi} \psi_\mu &= \xi^\nu\partial_\nu \psi_\mu
+\psi_\nu\partial_\mu \xi^\nu\,.
\end{align}
In the differential-form notation, these transformations take the compact form
\begin{align}
\delta_{\xi} \omega^{ab}=\mathcal{L}_{\xi}\omega^{ab}
&= \xi\lrcorner(d\,\omega^{ab})+d(\xi\lrcorner\,\omega^{ab})\,,\\
\delta_{\xi} e^{a}=\mathcal{L}_{\xi}e^{a}
&= \xi\lrcorner(d\,e^{a})+d(\xi\lrcorner\,e^{a})\,,\\
\delta_{\xi} \psi=\mathcal{L}_{\xi}\psi
&= \xi\lrcorner(d\,\psi)+d(\xi\lrcorner\,\psi)\,.
\end{align}
Later we will use the helpful purely bosonic identity $\mathcal{L}_{\xi}A^{IJ}=\xi\lrcorner F(A)^{IJ}+D^{A} (\xi\lrcorner A^{IJ})$ extended later to the supersymmetric version and its fermionic analog $\mathcal{L}_{\xi}\psi=\xi\lrcorner \mathcal{F}+\mathcal{D}^{A}(\xi\lrcorner \psi)$.

%%%%%%%%%%%%%%%%%%%%%%%%%%%%%%%%%%%%%%%%%%%%%%%%%%%%%%%%%%%%%%%%%%

\section{Charges of supergravity}

In the following sections, we generalize the results of \cite{Durka:2021ftc} to the case
of supergravity. We first consider the symplectic structure in the bulk and at the corner and then we construct the conserved charges associated with local symmetries of the action, Lorentz symmetry, supersymmetry, and general coordinate invariance. Although the translational symmetry is not, strictly speaking, a symmetry of the action we can still construct the associated charges that, as it will turn out, vanish. Finally, we will discuss the algebra of symmetries and the associated charges.

The symplectic potential density $\theta$ is evaluated from the variation of the action
\begin{equation}
16\pi\,\delta S=\int_{M}d\theta+\ldots
\end{equation}
where $"\ldots"$ denotes terms that vanish when the field equations are
satisfied. Since the symplectic potential density $\theta$ is a spacetime three-form it can be integrated over the three dimensional surface $\Sigma$ to give the symplectic potential $\Theta$,
\begin{align}
\Theta = \int_\Sigma \theta\,.
\end{align}
In the case of the action \eqref{Lagrangian_sugra_fermionic} we get
\begin{equation}
16\pi\,\Theta=\int_{\Sigma}\mathbb B\wedge\delta \mathbb A=\int_{\Sigma}B^{(s)}_{IJ}\wedge\delta A^{IJ}\,+4\kappa\,\bar
{\mathcal{B}}\wedge\delta\psi\,.\label{SusySymplPot}
\end{equation}
Using the decomposition of $A^{IJ}$ and taking into account that
super-torsion vanishes, we obtain
\begin{align}
16\pi\,\Theta & \thickapprox\frac{3}{G\Lambda}\int_{\Sigma}\left(
R^{ab}(\omega)-\frac{\Lambda}{3}\,e^{[a}\,\wedge e^{b]}-\frac{\kappa^2}{2}\bar{\psi}\,\gamma^{ab}\wedge \psi\right) \left(
\gamma\delta_{ab}^{cd}-\frac{1}{2}\,\epsilon^{cd}{}_{ab}\right)
\,\wedge\delta\omega_{cd}\nonumber\\
& +\frac{3}{G\Lambda}\int_{\Sigma}4\kappa^{2}\left( d\bar{\psi}-\frac{1}%
{4}\omega^{ab}\wedge\bar{\psi}\,\gamma_{ab}-\frac{1}{2\ell}e^{a}\wedge\bar{\psi}\,\gamma_{a}\right) \left( \gamma
{\mathchoice {\rm 1\mskip-4mu l} {\rm 1\mskip-4mu l} {\rm 1\mskip-4.5mu l} {\rm 1\mskip-5mu l}}%
-\,\gamma^{5}\,\right) \wedge\delta\psi\,,
\end{align}
where $"\thickapprox"$ denotes equalities holding up to the field equations, and as before $\kappa^{2}=\frac{4\pi G}{\ell}$ and $-\frac{1}{\ell^{2}%
}=\frac{\Lambda}{3}.$

Using the fact that super-torsion vanishes we can decompose $\Theta$ into the bulk $\Sigma$ contribution
\begin{align}
\Theta^{\Sigma} & \thickapprox-\frac{1}{16\pi G}\,\int_{\Sigma}\,e^{a}\wedge
e^{b}\left( \gamma\delta_{ab}^{cd}-\frac{1}{2}\,\epsilon^{cd}{}_{ab}\right)
\wedge\delta\omega_{cd}\nonumber\\
& +\frac{3}{16\pi G\Lambda}\int_{\Sigma}\left( \omega^{af}\wedge\omega
_{f}{}^{b}-\frac{\kappa^2}{2}\bar{\psi}\,\gamma^{ab}\wedge\psi\right) \left( \gamma\delta_{ab}^{cd}-\frac{1}{2}\,\epsilon^{cd}%
{}_{ab}\right) \,\wedge\delta\omega_{cd}\nonumber\\
& +\frac{3}{G\Lambda}\frac{4\pi G}{\ell}\int_{\Sigma}4\left( -\frac{1}%
{4}\omega^{ab}\wedge\bar{\psi}\,\gamma_{ab}-\frac{1}{2\ell}e^{a}\wedge\bar{\psi}\,\gamma_{a}\right) \left( \gamma
{\mathchoice {\rm 1\mskip-4mu l} {\rm 1\mskip-4mu l} {\rm 1\mskip-4.5mu l} {\rm 1\mskip-5mu l}}%
-\,\gamma^{5}\,\right) \wedge\delta\psi\,,
\end{align}
and
the corner $S=\partial\Sigma$ one
\begin{align}
\Theta^{S} & \thickapprox\frac{3}{16\pi G\Lambda}\,\int_{S}\omega{}%
^{ab}\left( \gamma\delta_{ab}^{cd}-\frac{1}{2}\,\epsilon^{cd}{}_{ab}\right)
\,\wedge\delta\omega_{cd} +\frac{3}{4\pi G\Lambda}\frac{4\pi G}{\ell}\int_{S}\bar{\psi}\left(
\gamma
{\mathchoice {\rm 1\mskip-4mu l} {\rm 1\mskip-4mu l} {\rm 1\mskip-4.5mu l} {\rm 1\mskip-5mu l}}%
-\,\gamma^{5}\,\right) \wedge\delta\psi\,.
\end{align}

%%%%%%%%%%%%%%%%%%%%%%%%%%%%%%%%%%%%%%%%%%%%%%%%%%%%%%%%%%%%%%%%%%

\section{Boundary charges}

We start with the BF-theory symplectic form derived from \eqref{SusySymplPot}
\begin{equation}
16\pi\,\Omega\equiv 16\pi\,\delta\Theta=\int_{\Sigma}\delta \mathbb B\wedge\delta \mathbb A=\int_{\Sigma}\delta B^{(s)}_{IJ}\wedge\delta A^{IJ}+4\kappa\,\delta
\bar{\mathcal{B}}\wedge\delta\psi\,.
\end{equation}
Given the symplectic form $\Omega$ and a symmetry $\delta_{\ast}\,,$ the
charge $\mathcal{H}$ is defined by the relation (see e.g., \cite{Harlow:2019yfa})
\begin{align}
\delta\mathcal{H}[\ast]&=-\delta_{\ast}\intprod\Omega=-\delta_{\ast}\intprod \frac{1}{16\pi}\int_{\Sigma}\delta \mathbb B\wedge\delta \mathbb A\nonumber\\
&=- \frac{1}{16\pi}\int_{\Sigma}\delta_{\ast} \mathbb B\wedge\delta \mathbb A+\delta \mathbb B\wedge\delta_{\ast} \mathbb A\,.
\end{align}

%%%%%%%%%%%%%%%%%%%%%%%%%%%%%%%%%%%%%%%%%%%%%%%%%%%%%%%%%%%%%%%%%%

\subsection{Gauge charges}

For the gauge parameter $\Upsilon$:
\begin{align}
\delta\mathcal{H}[\Upsilon]&=- \frac{1}{16\pi}\int_{\Sigma}\delta_{\Upsilon} \mathbb B\wedge\delta \mathbb A+\delta \mathbb B\wedge\delta_{\Upsilon} \mathbb A\,,
\end{align}
we use $\delta_\Upsilon \mathbb{A}_{\mu}=\partial_\mu \Upsilon
-i[\mathbb{A}_{\mu}, \Upsilon]\equiv \mathbb D^{\mathbb{A}}_\mu \Upsilon$ (see \eqref{c3-17}) and $\delta_\Upsilon \mathbb{B}=-i[ \Upsilon, \mathbb B]$
to get
\begin{align}
\delta\mathcal{H}[\Upsilon]&=- \frac{1}{16\pi}\int_{\Sigma}\delta_{\Upsilon} \mathbb B\wedge\delta \mathbb A+\delta \mathbb B\wedge \mathbb D^{\mathbb A} \Upsilon\nonumber\\
&=- \frac{1}{16\pi}\int_{\Sigma}-i[ \Upsilon, \mathbb B]\wedge\delta \mathbb A+d(\delta \mathbb B \Upsilon)-d(\delta \mathbb B) \Upsilon-i \delta \mathbb B \wedge [ \mathbb A,\Upsilon]\nonumber\\
&=- \frac{1}{16\pi}\int_{\Sigma} \delta ( d(\mathbb B \Upsilon))-\delta ((d\mathbb B) \Upsilon)-i[ \Upsilon, \mathbb B]\wedge\delta \mathbb A -i \delta \mathbb B \wedge [ \mathbb A,\Upsilon]\,,
\end{align}
which simplifies to
\begin{align}
\mathcal{H}[\Upsilon]&=- \frac{1}{16\pi}\int_{\Sigma} d( \mathbb B \Upsilon)-(\mathbb D^{\mathbb A} \mathbb B) \Upsilon\,.
\end{align}
The last term in this expression vanishes due to the field equations, and we see that only the corner contribution survives 
\begin{align}
\mathcal{H}[\Upsilon]&=- \frac{1}{16\pi}\int_{S} \mathbb B \Upsilon\,.
\end{align}

The complete list of charges for the Lorentz transformations, translations, and supersymmetry is given below:
\begin{align}
\mathcal{H}^{S}[\lambda]&=- \frac{1}{16\pi}\int_{S} B^{(s)}_{ab} \lambda^{ab}\,,\label{charge_lambda}\\
\mathcal{H}^{S}[\zeta]&=- \frac{2}{16\pi}\int_{S} B^{(s)}_{a} \zeta^{a}\,,\label{charge_zeta}\\
\mathcal{H}^{S}[\epsilon]&=- \frac{4}{16\pi}\int_{S} \bar{\mathcal{B}} \epsilon\,.\label{charge_epsilon}
\end{align}

%%%%%%%%%%%%%%%%%%%%%%%%%%%%%%%%%%%%%%%%%%%%%%%%%%%%%%%%%%%%%%%%%

\subsection{Diffeomorphism charges}

Let us now turn to the charges associated with diffeomorphisms, again
following \cite{Freidel:2020xyx}. The charge associated with diffeomorphisms
can be expressed as
\begin{equation}
-\mathcal{L}_{\xi}\intprod\Omega=\frac{1}{16\pi}\int_{\Sigma}\delta
B^{(s)}_{IJ}\wedge\mathcal{L}_{\xi}A^{IJ}-\mathcal{L}_{\xi}B^{(s)}_{IJ}\wedge\delta
A^{IJ}+4\kappa\delta\,\bar{\mathcal{B}}\wedge\mathcal{L}_{\xi}\psi
-4\kappa\mathcal{L}_{\xi}\bar{\mathcal{B}}\wedge\delta\psi\,,
\end{equation}
which (assuming that the parameter $\xi$ is field-independent) can be
rewritten
\begin{align}
-\mathcal{L}_{\xi}\intprod\,\Omega&=\frac{1}{16\pi}\delta\left( \int_{\Sigma
}B^{(s)}_{IJ}\wedge\mathcal{L}_{\xi}A^{IJ}\right) -\frac{1}{16\pi}\int_{S}%
\xi\intprod\left( B^{(s)}_{IJ}\wedge\mathcal{L}_{\xi}A^{IJ}\right) \nonumber\\
&+\frac{4\kappa}{16\pi}\delta\left( \int_{\Sigma
}\bar{\mathcal{B}}\wedge\mathcal{L}_{\xi}\psi\right) -\frac{4\kappa}{16\pi}\int_{S}%
\xi\intprod\left( \bar{\mathcal{B}}\wedge\mathcal{L}_{\xi}\psi\right) \,.
\end{align}

We assume that the only non-vanishing components of the vector field $\xi$ at
the corner $S$ are those tangent to $S$. Thus, the second terms in both lines of the expression above vanishes
identically\footnote{The generalization to the case of the vector fields $\xi$ requires employing the extended phase space formalism.\cite{Ciambelli:2021nmv,Ciambelli:2022cfr}. We will address this issue in the forthcoming publication.}. Therefore the diffeomorphism charge has the form
\begin{equation}
\mathcal{H}[\xi]=\frac{1}{16\pi}\int_{\Sigma}B^{(s)}_{IJ}\wedge\mathcal{L}_{\xi
}A^{IJ}+\frac{4\kappa}{16\pi} \int_{\Sigma
}\bar{\mathcal{B}}\wedge\mathcal{L}_{\xi}\psi \,.
\end{equation}

Applying the Cartan magic formula $\mathcal{L}_{\xi}(*)=\xi\lrcorner(d\,*)+d(\xi
\lrcorner\,*)$ to the connection \eqref{connection} we derive two identities
\begin{align}
\mathcal{L}_{\xi}\bar{\psi}=\xi\lrcorner \bar{\mathcal{F}}(A)+\mathcal{D}^{A}
(\xi\lrcorner \bar\psi)+\bar{\psi}\gamma_{IJ}(\xi\lrcorner A^{IJ}) \,,
\end{align}
and
\begin{align}
\mathcal{L}_{\xi}A^{IJ}=\xi\lrcorner F(A)^{(s)IJ}+D^{A}
(\xi\lrcorner A^{IJ})+\kappa^2 \bar{\psi}\gamma^{IJ} (\xi\lrcorner\psi)\,.
\end{align}
These identities together with the field equations $\mathbb{D}^{\mathbb{A}}\mathbb{B}=0$ lead to
\begin{align}
\mathcal{H}[\xi]&\thickapprox\frac{1}{16\pi}\int_{\Sigma}B^{(s)}_{IJ}\wedge
\xi\lrcorner F(A)^{(s)IJ}+d(B^{(s)}_{IJ}\xi\lrcorner A^{IJ})\nonumber\\
& +\frac{4\kappa}{16\pi}\int_{\Sigma} \bar{\mathcal{B}}\wedge
\xi\lrcorner \mathcal{F}+d(\bar{\mathcal{B}}\xi\lrcorner\psi)\,.
\end{align}
The diffeomorphism charge above decomposes into the corner component
\begin{equation}
\ \mathcal{H}^{S}[\xi]\thickapprox\frac{1}{16\pi}\int_{S}B^{(s)}_{IJ}\xi\lrcorner
A^{IJ}+4\kappa\,\bar{\mathcal{B}}\xi\lrcorner\psi\,,
\end{equation}
and the bulk part
\begin{equation}
\mathcal{H}^{\Sigma}[\xi]\thickapprox\frac{1}{16\pi}\int_{\Sigma}B^{(s)}_{IJ}%
\wedge\xi\lrcorner F(A)^{(s)IJ}+ 4\kappa\,\bar{\mathcal{B}}\wedge
\xi\lrcorner \mathcal{F}\,.
\end{equation}
Using field equations we find that 
\begin{equation}
\mathcal{H}^{\Sigma}[\xi]\thickapprox\frac{1}{2}\int_{\Sigma}\xi \lrcorner\, \text{Lagrangian}\,.
\end{equation}
In the case of $\xi$ being tangential to the $\Sigma$ the bulk contribution to the diffeomorphism charge vanishes; the case of general $\xi$ can be again handled using the extended phase space formalism.

%%%%%%%%%%%%%%%%%%%%%%%%%%%%%%%%%%%%%%%%%%%%%%%%%%%%%%%%%%%%%%%%%%

\section{Algebra of charges}

Let us now consider the charge algebra. We can compute the algebra of corner charges \eqref{charge_lambda}-\eqref{charge_epsilon} using the fact that by construction the charges generate infinitesimal transformations through the Poisson
bracket (see e.g., \cite{Harlow:2019yfa,Ciambelli:2022vot,Speziale:2025lkm})
\begin{equation}
\left\{\mathcal{H}^{S}[\Xi], (\ast)\right\} = \delta_{\Xi}(\ast)\,.
\end{equation}
The Poisson bracket of two charges can be expressed as a variation of the charge
\begin{equation}
\left\{ \mathcal{H}^{S}[\Xi_{1}], \mathcal{H}^{S}[\Xi_{2}]\right\}
=\frac12\left( \delta_{\Xi_{1}} \mathcal{H}^{S}[\Xi_{2}] -\delta_{\Xi_{2}}
\mathcal{H}^{S}[\Xi_{1}]\right) \,.
\end{equation}
We first consider the algebra of charges associated with local gauge symmetries, Lorentz, translational, and supersymmetry with the infinitesimal parameters ($\lambda^{ab}, \zeta^{a}, \epsilon_{\alpha}$). We assume that the gauge parameters are field-independent.

We use the following notation for the corner charges with labels referring to: Lorentz (L), translation (T) and supercharge (SUSY):
\begin{align}
\mathcal{H}_L[\lambda]&=- \frac{1}{16\pi}\int_{S} B^{(s)}_{ab} \lambda^{ab}\,,\\
\mathcal{H}_T[\zeta]&=- \frac{2}{16\pi}\int_{S} B^{(s)}_{a} \zeta^{a}\,,\\
\mathcal{H}_{SUSY}[\epsilon]& =- \frac{4}{16\pi}\int_{S} \bar{\mathcal{B}} \epsilon\,.
\end{align}

%%%%%%%%%%%%%%%%%%%%%%%%%%%%%%%%%%%%%%%%%%%%%%%%%%%%%%%%%%%%%%%%%%

\subsection{Lorentz with Lorentz: $\lambda$ with $\lambda$}

For two Lorentz charges we have
\begin{align}
\left\{
\mathcal{H}_L[\lambda_{1}],\mathcal{H}_L[\lambda_{2}]
\right\}
=
\frac12
\left(
\delta_{\lambda_{1}}\mathcal{H}_L[\lambda_{2}]
-
\delta_{\lambda_{2}}\mathcal{H}_L[\lambda_{1}]
\right)\,.
\end{align}
Using
\begin{equation}
\delta_{\lambda}B^{(s)}_{ab}
=
-\lambda_{a}{}^{c}B^{(s)}_{cb}
-\lambda_{b}{}^{c}B^{(s)}_{ac}\,,
\end{equation}
and
\begin{equation}
\lambda_{12}^{ab}
=
\lambda_{1}{}^{a}{}_{c}\lambda_{2}^{cb}
-
\lambda_{2}{}^{a}{}_{c}\lambda_{1}^{cb}\,,
\end{equation}
one finds
\begin{align}
\left\{
\mathcal{H}_L[\lambda_{1}],\mathcal{H}_L[\lambda_{2}]
\right\}
&=
-\frac{1}{16\pi}\int_{S}
B^{(s)}_{ab}\,\lambda_{12}^{ab}
=
\mathcal{H}_L\big[\lambda_{12}\big]\,.
\end{align}
Thus, the Poisson bracket algebra of Lorentz charges reproduces the Lorentz subalgebra.

%%%%%%%%%%%%%%%%%%%%%%%%%%%%%%%%%%%%%%%%%%%%%%%%%%%%%%%%%%%%%%%%%%

\subsection{Lorentz with translation: $\lambda$ with $\zeta$}

Next, consider the mixed Lorentz--translation bracket,
\begin{align}
\left\{
\mathcal{H}_L[\lambda_{1}],\mathcal{H}_T[\zeta_{2}]
\right\}
=
\frac12
\left(
\delta_{\lambda_{1}}\mathcal{H}_T[\zeta_{2}]
-
\delta_{\zeta_{2}}\mathcal{H}_L[\lambda_{1}]
\right)\,.
\end{align}
Using
\begin{equation}
\delta_{\lambda}B^{(s)}_{a}
=
-\lambda_{a}{}^{b}B^{(s)}_{b},
\qquad
\delta_{\zeta}B^{(s)}_{ab}
=
B^{(s)}_{a}\zeta_{b}
-
B^{(s)}_{b}\zeta_{a}\,,
\end{equation}
we obtain
\begin{align}
\delta_{\lambda_{1}}\mathcal{H}_T[\zeta_{2}]
-
\delta_{\zeta_{2}}\mathcal{H}_L[\lambda_{1}]
=
-\frac{4}{16\pi}\int_{S}
B^{(s)}_{a}\,
\lambda_{1\,b}{}^{a}\zeta_{2}^{b}\,.
\end{align}
Hence,
\begin{align}
\left\{
\mathcal{H}_L[\lambda_{1}],\mathcal{H}_T[\zeta_{2}]
\right\}
&=
-\frac{2}{16\pi}\int_{S}
B^{(s)}_{a}\,(\lambda_{1\,b}{}^{a}\zeta_{2}^{b})\,.
\end{align}
This is again of the form of a translation charge, with transformed parameter
\begin{equation}
\zeta_{12}^{a}
=
\lambda_{1\,b}{}^{a}\zeta_{2}^{b}\,,
\end{equation}
so that
\begin{align}
\left\{
\mathcal{H}_L[\lambda_{1}],\mathcal{H}_T[\zeta_{2}]
\right\}
=-\frac{2}{16\pi}\int_{S}
B^{(s)}_{a}\,\zeta_{12}^{a}=
\mathcal{H}_T\big[\zeta_{12}\big].
\end{align}

%%%%%%%%%%%%%%%%%%%%%%%%%%%%%%%%%%%%%%%%%%%%%%%%%%%%%%%%%%%%%%%%%%

\subsection{Lorentz with supersymmetry: $\lambda$ with $\epsilon$}

Finally, for the mixed Lorentz--supersymmetry bracket, we use
\begin{align}
\left\{
\mathcal{H}_L[\lambda_{1}],\mathcal{H}_{SUSY}[\epsilon_{2}]
\right\}
=
\frac12
\left(
\delta_{\lambda_{1}}\mathcal{H}_{SUSY}[\epsilon_{2}]
-
\delta_{\epsilon_{2}}\mathcal{H}_L[\lambda_{1}]
\right).
\end{align}
The relevant transformations are
\begin{equation}
\delta_{\lambda}\bar{\mathcal B}
=
\frac14\,\bar{\mathcal B}\gamma_{ab}\lambda^{ab},
\qquad
\delta_{\epsilon}B^{(s)}_{ab}
=
\bar\epsilon\,\gamma_{ab}\,\mathcal B\,.
\end{equation}
A short computation, using the Majorana flip identity, gives
\begin{align}
\delta_{\lambda_{1}}\mathcal{H}_{SUSY}[\epsilon_{2}]
-
\delta_{\epsilon_{2}}\mathcal{H}_L[\lambda_{1}]
=
-\frac{2}{16\pi}\int_{S}
\bar{\mathcal B}\,\gamma_{ab}\lambda_{1}^{ab}\epsilon_{2}\,,
\end{align}
and therefore
\begin{align}
\left\{
\mathcal{H}_L[\lambda_{1}],\mathcal{H}_{SUSY}[\epsilon_{2}]
\right\}
&=
-\frac{4}{16\pi}\int_{S}
\bar{\mathcal B}\,(\frac{1}{4}\gamma_{ab}\lambda_{1}^{ab}\epsilon_{2})\,.
\end{align}
This is again of the form of a supersymmetry charge, with transformed parameter
\begin{equation}
\epsilon_{12}
=
\frac14\,\lambda_{1}^{ab}\gamma_{ab}\epsilon_{2},
\end{equation}
so that
\begin{align}
\left\{
\mathcal{H}_L[\lambda_{1}],\mathcal{H}_{SUSY}[\epsilon_{2}]
\right\}
=
\mathcal{H}_{SUSY}\big[\epsilon_{12}\big].
\end{align}

%%%%%%%%%%%%%%%%%%%%%%%%%%%%%%%%%%%%%%%%%%%%%%%%%%%%%%%%%%%%%%%%%%

\subsection{Translation with translation: $\zeta$ with $\zeta$}

For two translational charges we have
\begin{align}
\left\{
\mathcal{H}_T[\zeta_{1}],\mathcal{H}_T[\zeta_{2}]
\right\}
=
\frac12
\left(
\delta_{\zeta_{1}}\mathcal{H}_T[\zeta_{2}]
-
\delta_{\zeta_{2}}\mathcal{H}_T[\zeta_{1}]
\right)\,,
\end{align}
and the required transformation is
\begin{equation}
\delta_{\zeta}B^{(s)}_{a}
=
B^{(s)}_{ab}\,\zeta^{b}.
\end{equation}
It follows that
\begin{align}
\delta_{\zeta_{1}}\mathcal{H}_T[\zeta_{2}]
-
\delta_{\zeta_{2}}\mathcal{H}_T[\zeta_{1}]
= -\frac{2}{16\pi}\int_{S} B^{(s)}_{ac} (\zeta_{1}^{c}\zeta_{2}^{a}-\zeta_{2}^{c}\zeta_{1}^{a})\,.
\end{align}
Hence
\begin{align}
\left\{
\mathcal{H}_T[\zeta_{1}],\mathcal{H}_T[\zeta_{2}]
\right\}
=
-\frac{1}{16\pi}\int_{S}
B^{(s)}_{ab}\,\left(-(\zeta_{1}^{a}\zeta_{2}^{b}-\zeta_{2}^{a}\zeta_{1}^{b})\right).
\end{align}
Defining
\begin{equation}
\lambda_{12}^{ab}:=-(\zeta_{1}^{a}\zeta_{2}^{b}-\zeta_{2}^{a}\zeta_{1}^{b})\,,
\end{equation}
allows us to write
\begin{align}
\left\{
\mathcal{H}_T[\zeta_{1}],\mathcal{H}_T[\zeta_{2}]
\right\}
=
\mathcal{H}_L\big[\lambda_{12}\big].
\end{align}

%%%%%%%%%%%%%%%%%%%%%%%%%%%%%%%%%%%%%%%%%%%%%%%%%%%%%%%%%%%%%%%%%%

\subsection{Translation with supersymmetry: $\zeta$ with $\epsilon$}

Let us consider the mixed translation--supersymmetry bracket
\begin{align}
\left\{
\mathcal{H}_T[\zeta_{1}],\mathcal{H}_{SUSY}[\epsilon_{2}]
\right\}
=
\frac12
\left(
\delta_{\zeta_{1}}\mathcal{H}_{SUSY}[\epsilon_{2}]
-
\delta_{\epsilon_{2}}\mathcal{H}_T[\zeta_{1}]
\right).
\end{align}
Using
\begin{equation}
\delta_{\zeta}\bar{\mathcal B}
=
\frac12\,\bar{\mathcal B}\gamma_{a}\zeta^{a},
\qquad
\delta_{\epsilon}B^{(s)}_{a}
=
-\bar\epsilon\,\gamma_{a}\mathcal B,
\end{equation}
we obtain
\begin{align}
\delta_{\zeta_{1}}\mathcal{H}_{SUSY}[\epsilon_{2}]
&=
-\frac{4}{16\pi}\int_{S}
\left(\delta_{\zeta_{1}}\bar{\mathcal B}\right)\epsilon_{2} = -\frac{2}{16\pi}\int_{S}
\bar{\mathcal B}\gamma_{a}\zeta_{1}^{a}\epsilon_{2}\,,
\end{align}
and
\begin{align}
\delta_{\epsilon_{2}}\mathcal{H}_T[\zeta_{1}]
&=
-\frac{2}{16\pi}\int_{S}
\left(\delta_{\epsilon_{2}}B^{(s)}_{a}\right)\zeta_{1}^{a} = \frac{2}{16\pi}\int_{S}
\bar\epsilon_{2}\gamma_{a}\mathcal B\,\zeta_{1}^{a}.
\end{align}
Using the Majorana flip identity
\begin{equation}
\bar\epsilon_{2}\gamma_{a}\mathcal B
=
-\bar{\mathcal B}\gamma_{a}\epsilon_{2}\,,
\end{equation}
we find
\begin{align}
\left\{
\mathcal{H}_T[\zeta_{1}],\mathcal{H}_{SUSY}[\epsilon_{2}]
\right\}
&=
-\frac{4}{16\pi}\int_{S}
\bar{\mathcal B}(\frac{1}{2}\gamma_{a}\epsilon_{2}\zeta_{1}^{a})\,.
\end{align}
This is again of the form of a supersymmetry charge. Defining
\begin{equation}
\epsilon_{12}
=
\frac12\,\zeta_{1}^{a}\gamma_{a}\epsilon_{2}\,,
\end{equation}
we can write
\begin{align}
\left\{
\mathcal{H}_T[\zeta_{1}],\mathcal{H}_{SUSY}[\epsilon_{2}]
\right\}
=
\mathcal{H}_{SUSY}\big[\epsilon_{12}\big].
\end{align}

%%%%%%%%%%%%%%%%%%%%%%%%%%%%%%%%%%%%%%%%%%%%%%%%%%%%%%%%%%%%%%%%%%

\subsection{Supersymmetry with supersymmetry: $\epsilon$ with $\epsilon$}

Let us now finally consider the bracket of two supersymmetry charges,
\begin{align}
\left\{
\mathcal{H}_{SUSY}[\epsilon_{1}],\mathcal{H}_{SUSY}[\epsilon_{2}]
\right\}
=
\frac12
\left(
\delta_{\epsilon_{1}}\mathcal{H}_{SUSY}[\epsilon_{2}]
-
\delta_{\epsilon_{2}}\mathcal{H}_{SUSY}[\epsilon_{1}]
\right).
\end{align}
Using
\begin{equation}
\delta_{\epsilon}\bar{\mathcal B}
=
-\frac14\,\bar\epsilon\,\gamma^{ab}B^{(s)}_{ab}
-\frac12\,\bar\epsilon\,\gamma^{a}B^{(s)}_{a},
\end{equation}
we obtain
\begin{align}
\delta_{\epsilon_{1}}\mathcal{H}_{SUSY}[\epsilon_{2}]
&=
\frac{1}{16\pi}\int_{S}
\left(\bar\epsilon_{1}\gamma^{ab}\epsilon_{2}\right)B^{(s)}_{ab}
+
\frac{2}{16\pi}\int_{S}
\left(\bar\epsilon_{1}\gamma^{a}\epsilon_{2}\right)B^{(s)}_{a},
\\
\delta_{\epsilon_{2}}\mathcal{H}_{SUSY}[\epsilon_{1}]
&=
\frac{1}{16\pi}\int_{S}
\left(\bar\epsilon_{2}\gamma^{ab}\epsilon_{1}\right)B^{(s)}_{ab}
+
\frac{2}{16\pi}\int_{S}
\left(\bar\epsilon_{2}\gamma^{a}\epsilon_{1}\right)B^{(s)}_{a}.
\end{align}
After using the standard Majorana identities for the bilinears, the bracket becomes
\begin{align}
\left\{
\mathcal{H}_{SUSY}[\epsilon_{1}],\mathcal{H}_{SUSY}[\epsilon_{2}]
\right\}
&=
-\frac{1}{16\pi}\int_{S}
B^{(s)}_{ab}\left(\bar\epsilon_{1}\gamma^{ab}\epsilon_{2}\right)
-\frac{2}{16\pi}\int_{S}
B^{(s)}_{a}\left(\bar\epsilon_{1}\gamma^{a}\epsilon_{2}\right).
\end{align}
Therefore
\begin{align}
\left\{
\mathcal{H}_{SUSY}[\epsilon_{1}],\mathcal{H}_{SUSY}[\epsilon_{2}]
\right\}
=
\mathcal{H}_L\big[\lambda_{12}\big]
+
\mathcal{H}_T\big[\zeta_{12}\big],
\end{align}
with composite parameters
\begin{equation}
\lambda_{12}^{ab}
=
\bar\epsilon_{1}\gamma^{ab}\epsilon_{2},
\qquad
\zeta_{12}^{a}
=
\bar\epsilon_{1}\gamma^{a}\epsilon_{2}\,.
\end{equation}

%%%%%%%%%%%%%%%%%%%%%%%%%%%%%%%%%%%%%%%%%%%%%%%%%%%%%%%%%%%%%%%%%%

\subsection{Summary of algebra of (gauge) charges and the on-shell reduction}

By the explicit calculation we thus convinced ourselves that the algebra of Poisson brackets of the Noether charges associated with local gauge symmetries represents the original supersymmetry algebra of $\OSp(1|4)$.

Collecting the results obtained above, the gauge part of the corner charge algebra can be summarized as
$$
\renewcommand{\arraystretch}{1.5}
\begin{array}{c|ccc}
 \{\cdot, \cdot\} & \mathcal{H}_L[\lambda_2] & \mathcal{H}_T[\zeta_2] & \mathcal{H}_{SUSY}[\epsilon_2] \\
\hline
 \mathcal{H}_L[\lambda_1]
 & \mathcal{H}_L[\lambda_{12}]
 & \mathcal{H}_T[\zeta_{12}]
 & \mathcal{H}_{SUSY}[\epsilon_{12}]
 \\
 \mathcal{H}_T[\zeta_1]
 & -
 & \mathcal{H}_L[\lambda_{12}]
 & \mathcal{H}_{SUSY}[\epsilon_{12}]
 \\
 \mathcal{H}_{SUSY}[\epsilon_1]
 & -
 & -
 & \mathcal{H}_L[\lambda_{12}] + \mathcal{H}_T[\zeta_{12}]
\end{array}
$$
where in each entry the composite parameters $\lambda_{12}$, $\zeta_{12}$ and $\epsilon_{12}$ are the ones defined in the corresponding subsections. The lower-triangular part is omitted since it is determined by antisymmetry of the Poisson bracket.

Before turning to diffeomorphisms, let us comment on the meaning of the above algebra on-shell. From the algebraic field equation for the translational component of the $\mathbb B$-field we have
\begin{equation}
B^{(s)}_{a}
=
\frac{1}{\beta}F^{(s)}_{a},
\end{equation}
which is the so-called super-torsion. The vanishing of the super-torsion due to field equations (as shown at the end of Appendix~\ref{AppendixC}) implies that the translational corner charge is weakly vanishing,
\begin{align}
\mathcal{H}_T[\zeta]
=
-\frac{2}{16\pi}\int_S B^{(s)}_{a} \zeta^a
=
-\frac{2}{16\pi\beta}\int_S F^{(s)}_{a} \zeta^a
\thickapprox 0\,.
\end{align}
Thus, while the translation sector is present in the formal gauge-charge algebra, it does not define an independent non-trivial corner charge on the physical shell. In particular, the translational contribution appearing in the supersymmetry bracket drops out weakly on-shell, and one finds
\begin{align}
\left\{
\mathcal{H}_{SUSY}[\epsilon_{1}],\mathcal{H}_{SUSY}[\epsilon_{2}]
\right\}
&=
\mathcal{H}_L[\lambda_{12}]
+
\mathcal{H}_T[\zeta_{12}]
\nonumber\\
&\thickapprox
\mathcal{H}_L[\lambda_{12}]\,.
\end{align}
It is therefore natural to quotient the weakly vanishing translation charge out of the algebra. The remaining non-trivial on-shell corner algebra is then generated by the Lorentz and supersymmetry charges alone and takes the form
$$
\renewcommand{\arraystretch}{1.5}
\begin{array}{c|cc}
 \{\cdot, \cdot\}_{\text{on-shell}} & \mathcal{H}_L[\lambda_2] & \mathcal{H}_{SUSY}[\epsilon_2] \\
\hline
 \mathcal{H}_L[\lambda_1]
 & \mathcal{H}_L[\lambda_{12}]
 & \mathcal{H}_{SUSY}[\epsilon_{12}]
 \\
 \mathcal{H}_{SUSY}[\epsilon_1]
 & -
 & \mathcal{H}_L[\lambda_{12}]
\end{array}
$$
which is the non-trivial internal gauge algebra that survives after imposing the equations of motion. In this sense, the translational part of the formal gauge algebra is pure gauge on-shell, whereas the Lorentz and supersymmetry charges remain as genuine corner charges.

As we will see in the next subsection, the brackets involving diffeomorphisms provide the complementary part of the full boundary algebra.

%%%%%%%%%%%%%%%%%%%%%%%%%%%%%%%%%%%%%%%%%%%%%%%%%%%%%%%%%%%%%%%%%%

\subsection{Algebra of diffeomorphism charges}

The diffeomorphism charge we found was
\begin{equation}
\mathcal{H}_D[\xi]
=
\frac{1}{16\pi}\int_{S}
\left(
B^{(s)}_{IJ} \xi\lrcorner A^{IJ}
+
4\kappa\,\bar{\mathcal B}\xi\lrcorner \psi
\right),
\end{equation}
where $\xi$ is tangent to the corner $S$. 

%%%%%%%%%%%%%%%%%%%%%%%%%%%%%%%%%%%%%%%%%%%%%%%%%%%%%%%%%%%%%%%%%%

\subsubsection{Diffeomorphism with diffeomorphism: $\xi$ with $\xi$}

We first compute the bracket of two diffeomorphism charges. Using
\begin{equation}
\{\mathcal H_D[\xi_1],(\ast)\}=\delta_{\xi_1}(\ast)=\mathcal L_{\xi_1}(\ast)\,,
\end{equation}
we obtain
\begin{align}
\{\mathcal H_D[\xi_1],\mathcal H_D[\xi_2]\}
&=
\delta_{\xi_1}\mathcal H_D[\xi_2]
\nonumber\\
&=
\frac{1}{16\pi}\int_S
\left(
\mathcal L_{\xi_1}B^{(s)}_{IJ}\xi_2\lrcorner A^{IJ}
+
B^{(s)}_{IJ} \xi_2\lrcorner \mathcal L_{\xi_1}A^{IJ}
\right)
\nonumber\\
&\qquad
+\frac{4}{16\pi}\int_S
\left(
\mathcal L_{\xi_1}\bar{\mathcal B} \xi_2\lrcorner \psi
+
\bar{\mathcal B} \xi_2\lrcorner \mathcal L_{\xi_1}\psi
\right).
\end{align}
Using the identity
\begin{equation}
\xi_2\lrcorner \mathcal L_{\xi_1}(*)
=
\mathcal L_{\xi_1}(\xi_2\lrcorner\,*)
-
[\xi_1,\xi_2]\lrcorner\,(*)\,,
\end{equation}
we can rewrite this as
\begin{align}
\{\mathcal H_D[\xi_1],\mathcal H_D[\xi_2]\}
&=
\frac{1}{16\pi}\int_S
\mathcal L_{\xi_1}\!\left(
B^{(s)}_{IJ} \xi_2\lrcorner A^{IJ}
+
4\,\bar{\mathcal B}\xi_2\lrcorner \psi
\right)
\nonumber\\
&\qquad
-\frac{1}{16\pi}\int_S
\left(
B^{(s)}_{IJ}[\xi_1,\xi_2]\lrcorner A^{IJ}
+
4\,\bar{\mathcal B} [\xi_1,\xi_2]\lrcorner \psi
\right).
\end{align}
The first term vanishes since $S$ is closed and $\xi_1$ is tangent to $S$. Therefore
\begin{align}
\{\mathcal H_D[\xi_1],\mathcal H_D[\xi_2]\}
&=
-\frac{1}{16\pi}\int_S
\left(
B^{(s)}_{IJ} [\xi_1,\xi_2]\lrcorner A^{IJ}
+
4\,\bar{\mathcal B}[\xi_1,\xi_2]\lrcorner \psi
\right)
\nonumber\\
&=
-\mathcal H_D[[\xi_1,\xi_2]]\,.
\end{align}
Thus, with the present sign convention for $\mathcal H_D$, the diffeomorphism charges form an anti-representation of the Lie algebra of vector fields tangent to the corner.

%%%%%%%%%%%%%%%%%%%%%%%%%%%%%%%%%%%%%%%%%%%%%%%%%%%%%%%%%%%%%%%%%%

\subsubsection{Lorentz with diffeomorphism: $\lambda$ with $\xi$}

We now compute the mixed Lorentz--diffeomorphism bracket. Using
\begin{equation}
\mathcal H_L[\lambda]
=
-\frac{1}{16\pi}\int_S B^{(s)}_{ab} \lambda^{ab},
\end{equation}
we find
\begin{align}
\{\mathcal H_D[\xi],\mathcal H_L[\lambda]\}
&=
\delta_\xi \mathcal H_L[\lambda]
=
-\frac{1}{16\pi}\int_S \mathcal L_\xi B^{(s)}_{ab} \lambda^{ab}.
\end{align}
Since
\begin{equation}
\mathcal L_\xi\!\left(B^{(s)}_{ab} \lambda^{ab}\right)
=
\mathcal L_\xi B^{(s)}_{ab}\lambda^{ab}
+
B^{(s)}_{ab} \mathcal L_\xi \lambda^{ab},
\end{equation}
and the integral of a total Lie derivative over $S$ vanishes, we get
\begin{align}
\{\mathcal H_D[\xi],\mathcal H_L[\lambda]\}
&=
\frac{1}{16\pi}\int_S B^{(s)}_{ab}\ \mathcal L_\xi \lambda^{ab}
=
-\mathcal H_L[\mathcal L_\xi \lambda].
\end{align}
By antisymmetry of the Poisson bracket,
\begin{equation}
\{\mathcal H_L[\lambda],\mathcal H_D[\xi]\}
=
\mathcal H_L[\mathcal L_\xi \lambda]\,.
\end{equation}

%%%%%%%%%%%%%%%%%%%%%%%%%%%%%%%%%%%%%%%%%%%%%%%%%%%%%%%%%%%%%%%%%%

\subsubsection{Translation with diffeomorphism: $\zeta$ with $\xi$}

For the translation charge
\begin{equation}
\mathcal H_T[\zeta]
=
-\frac{2}{16\pi}\int_S B^{(s)}_{a} \zeta^{a},
\end{equation}
one similarly finds
\begin{align}
\{\mathcal H_D[\xi],\mathcal H_T[\zeta]\}
&=
\delta_\xi \mathcal H_T[\zeta]
=
-\frac{2}{16\pi}\int_S \mathcal L_\xi B^{(s)}_{a} \zeta^{a}
\nonumber\\
&=
\frac{2}{16\pi}\int_S B^{(s)}_{a} \mathcal L_\xi \zeta^{a}
=
-\mathcal H_T[\mathcal L_\xi \zeta].
\end{align}
Therefore,
\begin{equation}
\{\mathcal H_T[\zeta],\mathcal H_D[\xi]\}
=
\mathcal H_T[\mathcal L_\xi \zeta]\,.
\end{equation}

%%%%%%%%%%%%%%%%%%%%%%%%%%%%%%%%%%%%%%%%%%%%%%%%%%%%%%%%%%%%%%%%%%

\subsubsection{Supersymmetry with diffeomorphism: $\epsilon$ with $\xi$}

Finally, for the supersymmetry charge
\begin{equation}
\mathcal H_{SUSY}[\epsilon]
=
-\frac{4}{16\pi}\int_S \bar{\mathcal B} \epsilon\,,
\end{equation}
we obtain
\begin{align}
\{\mathcal H_D[\xi],\mathcal H_{SUSY}[\epsilon]\}
&=
\delta_\xi \mathcal H_{SUSY}[\epsilon]
=
-\frac{4}{16\pi}\int_S \mathcal L_\xi \bar{\mathcal B} \space\epsilon\,.
\end{align}
Using again that the total Lie derivative integrates to zero on $S$, we get
\begin{align}
\{\mathcal H_D[\xi],\mathcal H_{SUSY}[\epsilon]\}
&=
\frac{4}{16\pi}\int_S \bar{\mathcal B} \mathcal L_\xi \epsilon
=
-\mathcal H_{SUSY}[\mathcal L_\xi \epsilon]\,.
\end{align}
Hence,
\begin{equation}
\{\mathcal H_{SUSY}[\epsilon],\mathcal H_D[\xi]\}
=
\mathcal H_{SUSY}[\mathcal L_\xi \epsilon]\,.
\end{equation}
Here $\mathcal L_\xi \epsilon$ denotes the Lie derivative of the spinorial parameter along the corner; more precisely, one should understand it as the corresponding spinorial Lie derivative on $S$.

%%%%%%%%%%%%%%%%%%%%%%%%%%%%%%%%%%%%%%%%%%%%%%%%%%%%%%%%%%%%%%%%%%

\subsubsection{Summary}

Collecting these results together with the internal gauge algebra derived above, we obtain
$$
\renewcommand{\arraystretch}{1.5}
\begin{array}{c|cccc}
 \{\cdot,\cdot\} & \mathcal{H}_L[\lambda_2] & \mathcal{H}_T[\zeta_2] & \mathcal{H}_{SUSY}[\epsilon_2] & \mathcal{H}_D[\xi_2] \\
\hline
 \mathcal{H}_L[\lambda_1]
 & \mathcal{H}_L[\lambda_{12}]
 & \mathcal{H}_T[\zeta_{12}]
 & \mathcal{H}_{SUSY}[\epsilon_{12}]
 & \mathcal{H}_L[\mathcal L_{\xi_2}\lambda_1]
 \\
 \mathcal{H}_T[\zeta_1]
 & -
 & \mathcal{H}_L[\lambda_{12}]
 & \mathcal{H}_{SUSY}[\epsilon_{12}]
 & \mathcal{H}_T[\mathcal L_{\xi_2}\zeta_1]
 \\
 \mathcal{H}_{SUSY}[\epsilon_1]
 & -
 & -
 & \mathcal{H}_L[\lambda_{12}] + \mathcal{H}_T[\zeta_{12}]
 & \mathcal{H}_{SUSY}[\mathcal L_{\xi_2}\epsilon_1]
 \\
 \mathcal{H}_D[\xi_1]
 & -
 & -
 & -
 & -\mathcal{H}_D[[\xi_1,\xi_2]]
\end{array}
$$
where, as before, the lower-triangular part is omitted since it is determined by antisymmetry.

On-shell, the translation charge weakly vanishes, and therefore all terms proportional to $\mathcal H_T$ drop out.

%%%%%%%%%%%%%%%%%%%%%%%%%%%%%%%%%%%%%%%%%%%%%%%%%%%%%%%%%%%%%%%%%%

\section{Conclusions}
In this work, we have derived and analyzed the conserved charges of $\mathcal{N}=1$ supergravity formulated as a constrained BF theory, placing particular emphasis on boundary and corner contributions. Using the covariant phase space approach, we identified the relevant symplectic structure and showed how conserved charges emerge from local gauge symmetries and diffeomorphisms.

A central result is that the algebra of boundary (corner) charges reproduces the expected superalgebra structure, confirming consistency with the underlying symmetry principles of supergravity. We further demonstrated that translational charges vanish on-shell as a consequence of the super-torsion constraint, and therefore do not contribute to the physical boundary symmetry algebra. The non-trivial on-shell algebra is thus generated by Lorentz and supersymmetry charges.

We also found that incorporating diffeomorphism charges completes the boundary symmetry structure, yielding a representation of the algebra of vector fields on the boundary. Altogether, these results highlight the role of boundary degrees of freedom in supergravity and offer a step toward extending the “corner symmetry” program to supersymmetric settings.

Among the many questions opened by this work, two appear particularly pressing. First, it would be interesting to apply the extended phase space formalism \cite{Ciambelli:2021nmv,Ciambelli:2022cfr} to generalize our analysis to arbitrary diffeomorphisms, beyond those tangential to the corner considered here. Second, one may explore whether the ambiguities in the definition of corner charges and the treatment of edge modes can be used to render the gauge translational charge non-vanishing. Both corner-orthogonal diffeomorphisms and gauge translational charges are physically significant, as they are related, for instance, to the energy and momentum of bulk configurations. Clarifying their interpretation, as well as the relation between them, remains an important direction for future work. These questions will be addressed in a forthcoming paper.

%%%%%%%%%%%%%%%%%%%%%%%%%%%%%%%%%%%%%%%%%%%%%%%%%%%%%%%%%%%%%%%%%%
\appendix
%%%%%%%%%%%%%%%%%%%%%%%%%%%%%%%%%%%%%%%%%%%%%%%%%%%%%%%%%%%%%%%%%%

\section{The $\OSp(1|4)$ superalgebra}\label{AppendixA}

We follow \cite{Nicolai:1984hb}, adapted to the signature conventions of \cite{Durka:2012wd}, to provide the $\OSp(1|4)$ superalgebra. The bosonic subalgebra of $\OSp(1|4)$ is $\SO(2,3)$, generated by $M_{IJ}$ with $I,J=0,1,2,3,4$, and satisfying
\begin{align}
[M_{IJ},M_{KL}]
&=
i\left(
\eta_{IL}M_{JK}
+\eta_{JK}M_{IL}
-\eta_{IK}M_{JL}
-\eta_{JL}M_{IK}
\right)\,,
\label{aa1}
\end{align}
where
\begin{align}
\eta_{IJ}=\mathrm{diag}(-,+,+,+,-)\,.
\end{align}
It is convenient to decompose the AdS generators into Lorentz generators $M_{ab}$ and translations according to
\begin{align}
M_{a4}:=P_a\,,
\qquad
a,b,c,d=0,1,2,3\,.
\end{align}
The bosonic algebra then takes the form
\begin{align}
[M_{ab},M_{cd}]
&=
i\left(
\eta_{ad}M_{bc}
+\eta_{bc}M_{ad}
-\eta_{ac}M_{bd}
-\eta_{bd}M_{ac}
\right)\,,
\label{aa2}
\\
[M_{ab},P_c]
&=
i\left(
\eta_{bc}P_a-\eta_{ac}P_b
\right)\,,
\label{aa3}
\\
[P_a,P_b]
&=
iM_{ab}\,.
\label{aa4}
\end{align}

The $\gamma$ matrices satisfy the Clifford algebra
\begin{equation}\label{aa4a}
\{\gamma^a,\gamma^b\}=2\eta^{ab}\,,
\qquad
\gamma_{ab}:=\frac{1}{2}[\gamma_a,\gamma_b]\,,
\end{equation}
with
\begin{equation}
\eta^{ab}=\mathrm{diag}(-,+,+,+)\,.
\end{equation}
A convenient explicit representation is
\begin{align}
\gamma^{5}
&=
\begin{pmatrix}
-i\sigma^2 & 0\\
0 & i\sigma^2
\end{pmatrix},
\qquad
\gamma^{0}
=
\begin{pmatrix}
0 & -i\sigma^2\\
-i\sigma^2 & 0
\end{pmatrix},
\\[1ex]
\gamma^{1}
&=
\begin{pmatrix}
\sigma^3 & 0\\
0 & \sigma^3
\end{pmatrix},
\qquad
\gamma^{2}
=
\begin{pmatrix}
0 & i\sigma^2\\
-i\sigma^2 & 0
\end{pmatrix},
\qquad
\gamma^{3}
=
\begin{pmatrix}
-\sigma^1 & 0\\
0 & -\sigma^1
\end{pmatrix}\,.
\label{c3-7}
\end{align}

Introducing
\begin{equation}\label{aa4b}
m_{a4}=\frac{1}{2}\,\gamma_a\,,
\qquad
m^{a4}=-\frac{1}{2}\,\gamma^a\,,
\qquad
m_{ab}=\frac{1}{4}[\gamma_a,\gamma_b]=\frac{1}{2}\gamma_{ab}\,,
\qquad
m^{ab}=\frac{1}{2}\gamma^{ab}\,,
\end{equation}
the action of the bosonic generators on the supercharges is
\begin{align}
[M_{IJ},Q_\alpha]
&=
i(m_{IJ})_\alpha{}^\beta\,Q_\beta\,,
\label{aa5}
\end{align}
or equivalently,
\begin{align}
[M_{ab},Q]
&=
\frac{i}{2}\gamma_{ab}Q\,,
\qquad
[P_a,Q]
=
\frac{i}{2}\gamma_aQ\,.
\label{aa6}
\end{align}

The supersymmetry anticommutator closes on the bosonic generators as
\begin{align}
\{Q_\alpha,Q_\beta\}
&=
-i(Cm^{IJ})_{\alpha\beta}M_{IJ}
\label{aa7}
\\
&=
-\frac{i}{2}(C\gamma^{ab})_{\alpha\beta}M_{ab}
+i(C\gamma^a)_{\alpha\beta}P_a\,.
\label{aa8}
\end{align}
Here $C$ is the charge-conjugation matrix. In the Majorana basis used in this paper one may choose $C=\gamma^2$.

%%%%%%%%%%%%%%%%%%%%%%%%%%%%%%%%%%%%%%%%%%%%%%%%%%%%%%%%%%%%%%%%%%%%%%%%%%%%%%

\section{Conventions}\label{AppendixB}

In this appendix we collect the conventions and identities used throughout the paper.

\subsection*{Differential forms}

Let $A$ be a $p$-form and $B$ a $q$-form. Then
\begin{align}
A\wedge B &= (-1)^{pq} B\wedge A\,,
\\
d(A\wedge B) &= dA\wedge B + (-1)^p A\wedge dB\,.
\end{align}
For the interior product with a vector field $\xi$ we use
\begin{align}
\xi\lrcorner(A\wedge B)
&=
(\xi\lrcorner A)\wedge B + (-1)^p A\wedge(\xi\lrcorner B)\,,
\end{align}
and Cartan's formula for the Lie derivative,
\begin{align}
\mathcal L_\xi &= \xi\lrcorner\, d + d(\xi\lrcorner\,\cdot)\,.
\end{align}

\subsection*{Levi-Civita symbols and tetrads}

We define
\begin{align}
\epsilon^{01234}=1\,,
\qquad
\epsilon^{0123}=1\,,
\qquad
\epsilon_{0123}=-1\,,
\end{align}
so that
\begin{align}
\epsilon^{abcd4}=\epsilon^{abcd}\,.
\end{align}
The generalized Kronecker delta is
\begin{align}
\delta^{ab}_{cd}=\delta^a_c\delta^b_d-\delta^a_d\delta^b_c\,,
\end{align}
and the standard contraction identity reads
\begin{align}
\epsilon^{abmn}\epsilon_{mncd}
=
-(4-2)!\,\delta^{ab}_{cd}
=
-2\,\delta^{ab}_{cd}\,.
\end{align}

The spacetime metric is constructed from the tetrad as
\begin{align}
g_{\mu\nu}=e^a{}_\mu e^b{}_\nu \eta_{ab}\,,
\end{align}
and we denote
\begin{align}
e\equiv \det(e^a{}_\mu)=\sqrt{-g}\,.
\end{align}
The volume density can be written as
\begin{align}
e
=
\frac{1}{4!}\,
\epsilon_{abcd}\,
\epsilon^{\mu\nu\rho\sigma}
\,e^a{}_\mu e^b{}_\nu e^c{}_\rho e^d{}_\sigma\,.
\end{align}

\subsection*{Gamma-matrix identities}

We define
\begin{align}
\gamma^5=\gamma^0\gamma^1\gamma^2\gamma^3\,,
\qquad
\gamma_5=\gamma_0\gamma_1\gamma_2\gamma_3\,,
\end{align}
so that
\begin{align}
\gamma^5
&=
-\frac{1}{4!}\epsilon_{abcd}\gamma^a\gamma^b\gamma^c\gamma^d\,,
\\
\gamma_5
&=
\frac{1}{4!}\epsilon^{abcd}\gamma_a\gamma_b\gamma_c\gamma_d\,,
\end{align}
and therefore
\begin{align}
\gamma^5=-\gamma_5\,.
\end{align}
It follows that
\begin{align}
(\gamma^0)^2=(\gamma^5)^2=-1\,,
\qquad
(\gamma^1)^2=(\gamma^2)^2=(\gamma^3)^2=1\,.
\end{align}

The identities used most often in the main text are
\begin{align}
\gamma^{ab}
&=
\frac{1}{2}\epsilon^{abcd}\gamma_{cd}\gamma_5\,,
\\
\gamma_{ab}
&=
-\frac{1}{2}\epsilon_{abcd}\gamma^{cd}\gamma^5\,,
\\
\gamma_c\gamma_{ab}
&=
\eta_{ca}\gamma_b-\eta_{cb}\gamma_a-\epsilon_{abcd}\gamma^d\gamma^5\,,
\\
\gamma_{ab}\gamma_c
&=
\eta_{cb}\gamma_a-\eta_{ca}\gamma_b-\epsilon_{abcd}\gamma^d\gamma^5\,.
\end{align}

\subsection*{Majorana bilinears and Fierz identities}

For Majorana spinors the following symmetry properties of bilinears are used repeatedly:
\begin{align}
\bar\psi\,\chi &= \bar\chi\,\psi\,,
&
\bar\psi\,\gamma_5\,\chi &= \bar\chi\,\gamma_5\,\psi\,,
&
\bar\psi\,\gamma_5\gamma_a\,\chi &= \bar\chi\,\gamma_5\gamma_a\,\psi\,.
\end{align}
In particular,
\begin{align}
\epsilon^{\mu\nu\rho\sigma}\,\bar\psi_\mu\,\Gamma\,\psi_\nu = 0\,,
\qquad
\text{for}
\qquad
\Gamma\in\{\mathbf{1},\gamma^5,\gamma^5\gamma^a\}\,.
\end{align}
A quartic identity used in simplifying the Lagrangian is
\begin{align}
(\bar\psi_\mu\,\Gamma\,\psi_\nu)(\bar\psi_\rho\,\Gamma\,\psi_\sigma)\,
\epsilon^{\mu\nu\rho\sigma}=0\,,
\end{align}
for the gamma-matrix structures $\Gamma$ relevant in the fermionic sector of the theory.

%%%%%%%%%%%%%%%%%%%%%%%%%%%%%%%%%%%%%%%%%%%%%%%%%%%%%%%%%%%%%%%%%%

\section{Supergravity Lagrangian and field equations}\label{AppendixC}

After substituting the explicit expressions for the $\mathbb{B}$ fields into the BF action, the resulting Lagrangian splits into fermionic and bosonic parts,
\begin{align}
4\pi \mathcal L^{sugra}_{fermionic}
&=
\frac{\alpha}{\alpha^2+\beta^2}\,
\bar{\mathcal{F}}\wedge
\left(
\frac{\beta\bbone-\alpha\gamma^5}{2\alpha}
\right)\mathcal{F}\,,
\\
16\pi \mathcal L^{sugra}_{bosonic}
&=
\frac{1}{\beta}\,
F^{(s)a4}\wedge F_{a4}^{(s)}
+\frac{1}{4}\,
M^{ab}{}_{cd}\,
F_{ab}^{(s)}\wedge F_{cd}^{(s)}\,,
\end{align}
where the parameters are related to the Immirzi parameter by
\begin{align}
\gamma=\frac{\beta}{\alpha}\,,
\end{align}
and
\begin{equation}
M^{ab}{}_{cd}
=
\frac{\alpha}{\alpha^2+\beta^2}
\left(
\gamma\,\delta^{ab}_{cd}-\epsilon^{ab}{}_{cd}
\right)\,.
\end{equation}
Thus, after adding the bosonic and fermionic contributions one finds that
\small
\begin{align}
\frac{16\pi G}{\ell^{2}}\mathcal{L}
&=
2\kappa^{2}
\left(
\mathcal{D}^{\omega}\bar{\psi}
-\frac{1}{2\ell}e^{a}\wedge\bar{\psi}\gamma_{a}
\right)
\wedge
\left(
\frac{1}{\gamma}\bbone-\gamma^{5}
\right)
\left(
\mathcal{D}^{\omega}\psi
+\frac{1}{2\ell}e^{b}\wedge\gamma_{b}\psi
\right)
\nonumber\\
&\quad
+\frac{1}{4}
\left(
R_{ab}
+\frac{1}{\ell^{2}}e_{a}\wedge e_{b}
-\frac{\kappa^{2}}{2}\bar{\psi}\wedge\gamma_{ab}\psi
\right)
\left(
\frac{1}{\gamma}\delta^{ab}_{cd}
-\epsilon^{ab}{}_{cd}
\right)
\wedge
\left(
R^{cd}
+\frac{1}{\ell^{2}}e^{c}\wedge e^{d}
-\frac{\kappa^{2}}{2}\bar{\psi}\wedge\gamma^{cd}\psi
\right)
\nonumber\\
&\quad
+\frac{1+\gamma^{2}}{\gamma}
\left(
\frac{1}{\ell}D^{\omega}e^{a}
+\frac{\kappa^{2}}{2}\bar{\psi}\wedge\gamma^{a}\psi
\right)
\eta_{ab}\eta_{44}
\wedge
\left(
\frac{1}{\ell}D^{\omega}e^{b}
+\frac{\kappa^{2}}{2}\bar{\psi}\wedge\gamma^{b}\psi
\right)\,.
\end{align}
\normalsize
At this stage, the terms proportional to $R^{ab}$ and bilinear in the spinors cancel between $\mathcal L^{sugra}_{fermionic}$ and $\mathcal L^{sugra}_{bosonic}$. Moreover, by means of the Fierz identities, the four-fermion terms vanish identically, and the remaining torsion-dependent terms simplify accordingly.

On the other hand, as we established in \eqref{Bfield_I}-\eqref{Bfield_II}, the relevant field equations of our supersymmetric BF theory are
\begin{align}
\mathcal{D}^A\bar{\mathcal{B}}+\kappa\,\bar{\psi}\wedge\gamma^{ab} B^{(s)}_{ab}+2\kappa\,\bar{\psi}\wedge\gamma^{a} B^{(s)}_{a}&=0\,,\label{Bfield1}\\
D^A B^{(s)a}+\kappa\,\bar{\psi}\wedge\gamma^{a}\mathcal{B}&=0\,,\label{Bfield2}\\
D^A B^{(s)ab}+\kappa\,\bar{\psi}\wedge\gamma^{ab}\mathcal{B}&=0\,,\label{Bfield3}
\end{align}
which reduce, respectively, to the Rarita--Schwinger equation, the Einstein equations, and the vanishing of super-torsion. The first two, namely \eqref{Bfield1} and \eqref{Bfield2}, read
\begin{align}
0
&=
\mathcal{D}^{A}\bar{\mathcal{F}}
\left(\frac{\beta}{\alpha}\bbone-\gamma^{5}\right)
+\kappa\,\bar{\psi}\wedge\gamma_{ab}\,
\frac{1}{2}
\left(
\frac{\beta}{\alpha}\delta_{cd}^{ab}
-\epsilon^{ab}{}_{cd}
\right)
\left(
F^{cd}-\frac{\kappa^{2}}{2}\bar{\psi}\wedge\gamma^{cd}\psi
\right)
\nonumber\\
&\qquad
+2\kappa\,\bar{\psi}\wedge\gamma_{a}
\frac{\alpha^{2}+\beta^{2}}{\alpha\beta}
\left(
F^{a}+\frac{\kappa^{2}}{2}\bar{\psi}\wedge\gamma^{a}\psi
\right)\,,
\end{align}
\begin{align}
0
&=
D^{\omega}
\left(
F^{a}+\frac{\kappa^{2}}{2}\bar{\psi}\wedge\gamma^{a}\psi
\right)
\frac{\alpha^{2}+\beta^{2}}{\beta}
+\kappa\,\bar{\psi}\wedge\gamma^{a}
\left(
\beta\bbone-\alpha\,\gamma^{5}
\right)\mathcal{F}
\nonumber\\
&\qquad
-\frac{1}{\ell}\,
e_{b}\wedge
\left(
F^{cd}-\frac{\kappa^{2}}{2}\bar{\psi}\wedge\gamma^{cd}\psi
\right)
\frac{1}{2}
\left(
\beta\delta_{cd}^{ab}
-\alpha\,\epsilon^{ab}{}_{cd}
\right)\,,
\end{align}
whereas equation \eqref{Bfield3} reads
\begin{align}
0
&=
\frac{1}{2}
\left(
\frac{\beta}{\alpha}\delta_{cd}^{ab}
-\epsilon^{ab}{}_{cd}
\right)
D^\omega F^{(s)cd}
+\kappa\,\bar{\psi}\wedge\gamma^{ab}
\left(
\frac{\beta}{\alpha}\bbone-\gamma^{5}
\right)\mathcal{F}
\nonumber\\
&\qquad
+\frac{1}{\ell}\frac{\alpha^{2}+\beta^{2}}{\alpha\beta}
\left(
e^{a}\wedge F^{(s)b}-e^{b}\wedge F^{(s)a}
\right)\,.
\end{align}
Next, to see the true meaning of the last equation, we use the gamma-matrix identity
\begin{align}
\gamma^{ab}\gamma^5
=
\frac{1}{2}\epsilon^{ab}{}_{cd}\gamma^{cd}\,,
\end{align}
and we find
\begin{align}
\kappa\,\bar{\psi}\wedge\gamma^{ab}
\left(
\frac{\beta}{\alpha}\bbone-\gamma^{5}
\right)\mathcal{F}
&=
\frac{\beta}{\alpha}\,
\kappa\,\bar{\psi}\wedge\gamma^{ab}\mathcal{F}
-\kappa\,\bar{\psi}\wedge\gamma^{ab}\gamma^5\mathcal{F}
\nonumber\\
&=
\frac{\beta}{\alpha}\,
\kappa\,\bar{\psi}\wedge\gamma^{ab}\mathcal{F}
-\frac{1}{2}\epsilon^{ab}{}_{cd}\,
\kappa\,\bar{\psi}\wedge\gamma^{cd}\mathcal{F}
\nonumber\\
&=
\frac{1}{2}
\left(
\frac{\beta}{\alpha}\delta^{ab}_{cd}
-\epsilon^{ab}{}_{cd}
\right)
\kappa\,\bar{\psi}\wedge\gamma^{cd}\mathcal{F}\,.
\end{align}
Therefore the first two terms combine into
\begin{align}
0
&=
\frac{1}{2}
\left(
\frac{\beta}{\alpha}\delta_{cd}^{ab}
-\epsilon^{ab}{}_{cd}
\right)
\left(
D^\omega F^{(s)cd}
+\kappa\,\bar{\psi}\wedge\gamma^{cd}\mathcal{F}
\right)
\nonumber\\
&\qquad
+\frac{1}{\ell}\frac{\alpha^{2}+\beta^{2}}{\alpha\beta}
\left(
e^{a}\wedge F^{(s)b}-e^{b}\wedge F^{(s)a}
\right)\,.
\end{align}
Using the Bianchi identity \eqref{genBianchi_1decomposition_explicite1} brings us
\begin{align}
0
&=
-\frac{1}{2\ell}
\left(
\frac{\beta}{\alpha}\delta_{cd}^{ab}
-\epsilon^{ab}{}_{cd}
\right)
\left(
e^c\wedge F^{(s)d}-e^d\wedge F^{(s)c}
\right)
\nonumber\\
&\qquad
+\frac{1}{\ell}\frac{\alpha^{2}+\beta^{2}}{\alpha\beta}
\left(
e^{a}\wedge F^{(s)b}-e^{b}\wedge F^{(s)a}
\right)\,.
\end{align}
Since
\begin{align}
\frac{1}{2}\delta^{ab}_{cd}
\left(
e^c\wedge F^{(s)d}-e^d\wedge F^{(s)c}
\right)
&=
e^a\wedge F^{(s)b}-e^b\wedge F^{(s)a}\,,
\\
\frac{1}{2}\epsilon^{ab}{}_{cd}
\left(
e^c\wedge F^{(s)d}-e^d\wedge F^{(s)c}
\right)
&=
\epsilon^{ab}{}_{cd}\,e^c\wedge F^{(s)d}\,,
\end{align}
and
\begin{align}
\frac{\alpha^{2}+\beta^{2}}{\alpha\beta}-\frac{\beta}{\alpha}
=
\frac{\alpha}{\beta}
=
\frac{1}{\gamma}\,,
\end{align}
the above equation becomes
\begin{align}\label{vanishingsupertorsion}
\left(
\frac{1}{\gamma}\delta^{ab}_{cd}
+\epsilon^{ab}{}_{cd}
\right)
e^c\wedge F^{(s)d}
=
0\,.
\end{align}
For finite real $\gamma$, the operator in parentheses is invertible on antisymmetric pairs of internal indices, and therefore
\begin{align}
e^a\wedge F^{(s)b}-e^b\wedge F^{(s)a}=0\,.
\end{align}
Assuming a nondegenerate tetrad, this implies
\begin{align}
F^{(s)a}=0\,,
\end{align}
that is, the super-torsion defined in \eqref{super-torsion} vanishes.

%%%%%%%%%%%%%%%%%%%%%%%%%%%%%%%%%%%%%%%%%%%%%%%%%%%%%%%%%%%%%%%%%%

\bibliographystyle{uiuchept}
\bibliography{Bibliography}

\end{document}